\documentclass[showpacs,preprintnumbers,twocolumn,amsmath,amssymb,
groupedaddress,superscriptaddress]{revtex4-1}
\usepackage{graphicx}
\usepackage{dcolumn}
\usepackage{bm}

\begin{document}

\title{Microscopic analysis of $^{11}$Li elastic scattering on protons and\\
breakup processes within $^{9}$Li+$2n$ cluster model}

\author{V.~K.~Lukyanov}
\affiliation{Joint Institute for Nuclear Research, Dubna 141980,
Russia}

\author{D.~N.~Kadrev}
\affiliation{Institute for Nuclear Research and Nuclear Energy,
Bulgarian Academy of Sciences, Sofia 1784, Bulgaria}

\author{E.~V.~Zemlyanaya}
\affiliation{Joint Institute for Nuclear Research, Dubna 141980,
Russia}

\author{A.~N.~Antonov}
\affiliation{Institute for Nuclear Research and Nuclear Energy,
Bulgarian Academy of Sciences, Sofia 1784, Bulgaria}

\author{K.~V.~Lukyanov}
\affiliation{Joint Institute for Nuclear Research, Dubna 141980,
Russia}

\author{M.~K.~Gaidarov}
\affiliation{Institute for Nuclear Research and Nuclear Energy,
Bulgarian Academy of Sciences, Sofia 1784, Bulgaria}

\author{K.~Spasova}
\affiliation{Institute for Nuclear Research and Nuclear Energy,
Bulgarian Academy of Sciences, Sofia 1784, Bulgaria}
\affiliation{University "Ep.~K. Preslavski",
Shumen 9712, Bulgaria}

\begin{abstract}
In the paper, the results of analysis of elastic scattering and
breakup processes in interactions of the $^{11}$Li nucleus with
protons are presented. The hybrid model of the microscopic optical
potential (OP) is applied. This OP includes the single-folding
real part, while its imaginary part is derived within the
high-energy approximation (HEA) theory. For the $^{11}$Li$+p$
elastic scattering, the microscopic large-scale shell model (LSSM)
density of $^{11}$Li is used. The depths of the real and imaginary
parts of OP are fitted to the elastic scattering data at 62, 68.4,
and 75 MeV/nucleon, being simultaneously adjusted to reproduce the
true energy dependence of the corresponding volume integrals. The
role of the spin-orbit potential is studied and predictions for
the total reaction cross sections are made. Also, the cluster
model, in which $^{11}$Li consists of $2n$-halo and the $^{9}$Li
core having its own LSSM form of density, is adopted. The
respective microscopic proton-cluster OP's are calculated and
folded with the density probability of the relative motion of both
clusters to get the whole $^{11}$Li$+p$ optical potential. The
breakup cross sections of $^{11}$Li at 62 MeV/nucleon and momentum
distributions of the cluster fragments are calculated. An analysis
of the single-particle density of $^{11}$Li within the same
cluster model accounting for the possible geometric forms of the
halo-cluster density distribution is performed.
\end{abstract}


\maketitle

\section{Introduction
\label{s:intro}}

Recent experiments with radioactive ion beams have opened a new
era in nuclear physics by providing the possibility to study the
light nuclei far from stability. Indeed, the availability of the
radioactive ion beams favored the discovery of halo nuclei
\cite{Tanihata85a}. A typical example is the neutron halo in the
nucleus $^{11}$Li, revealed as a consequence of its very large
interaction radius, deduced from the measured interaction cross
sections of $^{11}$Li with various target nuclei
\cite{Tanihata85b,Tanihata88,Mittig87}. The halo of the nucleus
extends its matter distribution to a large radius. A hypothesis
based on the early data \cite{Tanihata85b} about the important
role played by the neutron pairing for the stability of nuclei
near the drip line is suggested in Refs.~\cite{Hansen87,Migdal73}
and, in particular, the direct link of the matter radius to the
$2n$ weak binding in $^{11}$Li is claimed to be attributed to its
configuration as a $^{9}$Li core coupled to a di-neutron.

The experiments that give evidences of the existence of a halo in this nucleus are related not only to measurements of the total reaction cross section for
$^{11}$Li projectiles but also to the momentum distributions of the $^{9}$Li or neutron fragments following the breakup of $^{11}$Li at high energies
\cite{Kobayashi88,Bertsch89,Anne90,Esbensen91}, e.g. the process $^{11}$Li+$^{12}$C at $E=800$ MeV/nucleon in Ref.~\cite{Kobayashi88}. Here we will mention also the
experiments at lower energies $E=60$ MeV/nucleon of scattering of $^{11}$Li on $^{9}$Be, $^{93}$Nb and $^{181}$Ta in \cite{Orr92} and of $^{11}$Li on a wide range
of nuclei from $^{9}$Be to $^{238}$U in \cite{Orr95}. It was shown that the momentum distribution of the breakup fragments has a narrow peak, much narrower than
that observed in the fragmentation of well bound nuclei. This property has been interpreted (e.g.,
\cite{Baye2010,Barranco96,Hencken96,Bertulani2004,Bertulani92,Ershov2004,Bertulani2002}) to be related to the very large extension of the wave function, as compared
to that of the core nucleus, leading to the existence of the nuclear halo. As pointed out in Ref.~\cite{Bertulani92}, the longitudinal component of the momentum
(taken along the beam or $z$ direction) gives the most accurate information on the intrinsic properties of the halo and is insensitive to details of the collision
and the size of the target.

The differential cross sections for small-angle proton elastic
scattering on Li isotopes at energies near 700 MeV/nucleon were
measured in inverse kinematics with secondary nuclear beams at GSI
(Darmstadt) \cite{Dobrovolsky2006}. They have been analyzed using
the Glauber theory and information on the nuclear matter density
distributions has been extracted. It was supposed that the two
valence neutrons in $^{11}$Li, which form the halo, could move in
a wide region far from the $^{9}$Li core that is related to the
small two-neutron separation energy ($\sim 0.3$ MeV).

The idea of existence of two-neutron halo in $^{11}$Li was
experimentally verified in measurements and studies of
differential cross sections of the $^{11}$Li$+p$ elastic
scattering in the energy range 60--75 MeV/nucleon
\cite{Moon92,Korsh97c,Korsh96}. The data analysis at 62
MeV/nucleon \cite{Moon92} showed that the adjusted
phenomenological Woods-Saxon (WS) potential has a shallow real
part and an imaginary part with a long tail. In
Refs.~\cite{Korsh97c,Korsh96} the data at 65--75 MeV/nucleon were
analyzed using the parameter free cluster-orbital shell-model
approximation (COSMA) \cite{Zhukov93} and a conclusion was drawn
that the $^{11}$Li$+p$ scattering is mainly determined by
scattering on the $^{9}$Li core. In various works (e.g.,
Refs.~\cite{Suzuki93,Kohno93,Chaudhuri94,Kanungo97,Kim2001}) the
calculations of the $^{11}$Li$+p$ differential cross sections in
the energy range $E<100$ MeV/nucleon differ between themselves by
the assumptions how the $^{11}$Li$+p$ optical potential to be
constructed. Most of them use the simple folding approach to the
real part of OP (ReOP) without accounting for the exchange terms
and with introducing different forms of effective nucleon-nucleon
(NN) interactions. To calculate the folding potentials, the
constituent $^{9}$Li+$2n$ cluster model was usually employed, in
which the $^{11}$Li density has two separated parts taken in
explicit forms. Various suggestions were made for the imaginary
part of OP (ImOP) like WS and Gaussian forms or calculated within
the $t$-matrix method. Then, the cross sections were computed
numerically by using the eikonal approximation or starting with
the Glauber multiple scattering theory. The more complicated model
of $^{11}$Li treated as a $^{9}$Li+n+n three-body system was
developed in Ref.~\cite{Crespo96}, where the effects of the halo
distribution in $^{11}$Li in correspondence to different parts of
the three-body wave function are manifested in the elastic cross
section.

Generally, here we would like to outline the advantages of the
microscopic analyses using the coordinate-space $g$-matrix folding
method (e.g., Ref.~\cite{Amos2005}), as well as works (e.g.,
Ref.~\cite{Avrigeanu2000}), where the ReOP is microscopically
calculated  using effective NN interactions within a folding
approach \cite{Satchler79,Khoa1993,Khoa2000,Khoa97} and including
also the exchange terms in it. In the recent works
\cite{Hassan2009,Farag2012} the $^{11}$Li$+p$ elastic scattering
cross sections were analyzed using folding procedure and effective
NN forces to calculate the real OP taking into account only its
direct part but not the exchange one. In Ref.~\cite{Hassan2009}
the volume ImOP was taken either in a WS form or in the form of
the direct folded ReOP and in Ref.~\cite{Farag2012} an application
of the microscopic OP \cite{Lukyanov2004a,Shukla2003} developed on
the base of the HEA theory \cite{Glauber,Sitenko} was also made.
To this end phenomenological densities (Gaussian-types and COSMA)
have been used in the calculations \cite{Hassan2009} and the LSSM
densities of $^{9,11}$Li \cite{Karataglidis97} in
Ref.~\cite{Farag2012}, as well.

The aims of our work can be presented as follows. First, we study elastic scattering cross section for $^{11}$Li$+p$ at three incident energies ($E<100$
MeV/nucleon) using microscopically calculated OP's within the hybrid model \cite{Lukyanov2004a}. The ReOP includes the direct and exchange terms and the ImOP is
based on the HEA. We follow our previous works \cite{Lukyanov2007,Lukyanov2009,Lukyanov2010}, where this model was applied to elastic scattering of exotic nuclei
$^{6,8}$He with use of their LSSM densities, and thus avoiding an adjustment of free parameters. As in Ref.~\cite{Lukyanov2009}, we pay attention to the ambiguity
problem when fitting the coefficients $N$'s that renormalize the strengths of different parts of OP. This ambiguity is minimized in Ref.~\cite{Lukyanov2012} by
testing the condition that the true energy dependence of the volume integrals must fulfill. Second, in addition to the analysis of elastic scattering cross
sections, we estimate other characteristics of the reaction mechanism such as the $^{11}$Li total reaction and breakup cross sections. The theoretical scheme used
in this second part of the work is based on the procedure from the first part to calculate microscopically the potentials necessary for the evaluation of the other
quantities within the model. The calculations are performed by using the $^{11}$Li+$p$ OP constructed as a sum of the microscopically calculated OP of $^{9}$Li+$p$
and the ($2n$-halo)+$p$ potential folded with a density probability of the relative motion of clusters. For a more consistent description of the halo structure of
$^{11}$Li we calculate the fragment momentum distributions from $^{11}$Li$+p$ reaction at 62 MeV/nucleon within the same breakup reaction model and present
predictions for them. Finally, we give results for the single-particle density distribution of $^{11}$Li within the true two-cluster model considering the relative
motion of clusters ($^{9}$Li+$h$) that is ensured by the respective wave function and make a comparison with other calculations.

The structure of this article is the following. The theoretical scheme to calculate microscopically the real and imaginary parts of the OP and the spin-orbit term,
as well as the results of the calculations of the elastic scattering of $^{11}$Li on protons and the discussion are given in Sec.~II. The next Sec.~III contains the
basic expressions to estimate the $^{11}$Li breakup and to calculate the momentum distributions of its products. The same Section contains the results of the total
breakup cross sections, the momentum distributions of clusters and the single-particle density distribution of $^{11}$Li calculated within the breakup model of
$^{11}$Li. The summary and conclusions of the work are given in Sec.~IV.

\section{Elastic scattering of $^{11}$Li on protons at $E<100$
MeV/nucleon}

\subsection{Microscopic ReOP}

The optical potential used in our calculations has the form
\begin{equation}
U_{opt}=V^{F}(r)+iW(r).
\label{eq:0}
\end{equation}
In Sec.~IIC we add also a spin-orbit term to $U_{opt}$ from
Eq.~(\ref{eq:0}).

The real part of the nucleon-nucleus OP is assumed to be a result
of a folding of the nuclear density and of the effective NN
potential and involves the direct and exchange parts (e.g.,
Refs.~\cite{Satchler79,Khoa1993,Khoa2000}, see also
\cite{Lukyanov2007,Lukyanov2009}):
\begin{equation}
V^{F}(r)= V^{D}(r)+V^{EX}(r).
\label{eq:1}
\end{equation}
The direct part $V^{D}(r)$ is composed by the isoscalar (IS) and
isovector (IV) contributions:
\begin{equation}
V^{D}_{IS}(r)=\int \rho_2({\bf r}_2)g(E)F(\rho_2)v_{00}^D(s)d{\bf
r}_2,
\label{eq:2}
\end{equation}
\begin{equation}
V^{D}_{IV}(r)=\int \delta\rho_2({\bf
r}_2)g(E)F(\rho_2)v_{01}^D(s)d{\bf r}_2
\label{eq:3}
\end{equation}
with ${\bf s}={\bf r}+{\bf r}_2$, and
\begin{equation}
\rho_2({\bf r}_2)=\rho_{2,p}({\bf r}_{2,p})+\rho_{2,n}({\bf
r}_{2,n}),
\label{eq:4}
\end{equation}
\begin{equation}
\delta\rho_2({\bf r}_2)=\rho_{2,p}({\bf r}_{2,p})-\rho_{2,n}({\bf
r}_{2,n}).
\label{eq:5}
\end{equation}
In Eqs.~(\ref{eq:4}) and (\ref{eq:5}) $\rho_{2,p}({\bf r}_{2,p})$
and $\rho_{2,n}({\bf r}_{2,n})$ are the proton and neutron
densities of the target nucleus. The expressions for the energy
and density dependence of the effective NN interaction  (the
formulae for $g(E)$ and $F(\rho)$) are given e.g., in
Ref.~\cite{Lukyanov2009}. For the NN potentials $v_{00}^D$ and
$v_{01}^D$ we use the expression from Ref.~\cite{Khoa2000} for the
CDM3Y6 type of the effective interaction based on the Paris NN
potential. The isoscalar part of the exchange contribution to the
ReOP has the form:
\begin{eqnarray}
V^{EX}_{IS}(r)&=&g(E)\int \rho_2({\bf r}_2, {\bf r}_2-{\bf s})
F\left(\rho_2({\bf r}_2-{\bf s}/2)\right ) \nonumber
\\ & \times & v_{00}^{EX}(s) j_0(k(r)s)d{\bf r}_2,
\label{eq:8}
\end{eqnarray}
$\rho_{2}$ being the one-body density matrix. It is shown in Ref.~\cite{Lukyanov2007} how the isovector part of the exchange ReOP can be obtained. Here we would
like to emphasize the general importance of the account for the exchange part of the OP. As shown on different examples in Ref.~\cite{Khoa2000}, the exchange
effects lead, for instance, to a particular energy dependence of the total potential, to different signs of the direct and exchange inelastic form factors and
others, so they should be treated as accurately as possible.

The LSSM proton and neutron densities used in our work for $^{11}$Li are calculated in a complex 2$\hbar\omega$ shell-model space using the WS basis of
single-particle wave functions with exponential asymptotic behavior \cite{Karataglidis97}, which is in principle the realistic one. Here we would like to discuss
this point. In many works, to simplify the analytical studies and calculations one uses basic functions and densities with Gaussian asymptotics of the type
$\exp(-ar^{2})$, while it has to be exponential one $\exp(-br)/r$, where the parameter $b$ is related to the bound energy of the particle in the upper shell. This
difference can affect the results for the cross sections in the region of relatively large angles of scattering. This point was one of the reasons the LSSM
densities \cite{Karataglidis97} for $^{9,11}$Li to be used in our work.

\subsection{Optical potential within the high-energy approximation}

In the present work we use the hybrid model of OP
\cite{Lukyanov2004a}, in which its imaginary part was derived
within the HEA theory \cite{Glauber,Sitenko}, while the real part
is obtained as prescribed by the folding procedure from Sec.~IIA.
The cross sections are calculated by means of the DWUCK4 code
\cite{DWUCK} for solving the Schr\"{o}dinger equation. To obtain
the HEA OP one can use the definition of the eikonal phase as an
integral of the nucleon-nucleus potential over the trajectory of
the straight-line propagation, and has to compare it with the
corresponding Glauber expression for the phase in the optical
limit approximation. In this way, the HEA OP is obtained as a
folding of the form factors of the nuclear density and the NN
amplitude $f_{NN}(q)$ \cite{Lukyanov2004a,Shukla2003}:
\begin{eqnarray}
U^H_{opt}&=&V^H+iW^H=-{\hbar v\over
(2\pi)^2}(\bar\alpha_{NN}+i)\bar\sigma_{NN}\nonumber \\
& \times & \int_0^\infty dq q^2 j_0(qr) \rho_2(q) f_{NN}(q).
\label{eq:14}
\end{eqnarray}
In Eq.~(\ref{eq:14}) $\bar\sigma_{NN}$ and $\bar\alpha_{NN}$ are,
respectively, the NN total scattering cross section and the ratio
of the real to imaginary part of the forward NN scattering
amplitude, both averaged over the isospin of the nucleus. These
two quantities have been parametrized in
\cite{Shukla2001,Charagi92} as functions of energies up to 1 GeV.
The values of $\bar\sigma_{NN}$ and $\bar\alpha_{NN}$ can also
account for the in-medium effect by a factor from
Ref.~\cite{Xiangzhow98}.

\subsection{The spin-orbit term}

The expression for the spin-orbit contribution to the OP used in
our work is added to the right-hand side of Eq.~(\ref{eq:0}) and
has the form:
\begin{equation}
V_{LS}(r)=2\lambda_{\pi}^{2}\left[V_{0}\frac{1}{r}\frac{df_{R}(r)}{dr}
+iW_{0} \frac{1}{r}\frac{df_{I}(r)}{dr}\right]({\bf l}\cdot{\bf s}),
\label{eq:15}
\end{equation}
where $\lambda_{\pi}^{2}$=2 fm$^{2}$ is the squared pion Compton
wavelength, $V_{0}$ and $W_{0}$ are the real and imaginary parts
of the microscopic OP at $r$=0. In our work, in Eq.~(\ref{eq:15})
the functions $f_{R}(r)$ and $f_{I}(r)$ are taken as WS forms
$f(r,R_{R},a_{R})$ and $f(r,R_{I},a_{I})$ with the half-radius
$R_{R}(R_{I})$ and diffuseness $a_{R}(a_{I})$ parameters obtained
by the best fit of the WS potential to the microscopically
calculated real $V(r)$ and imaginary $W(r)$ parts of the OP.

\subsection{Results of calculations of $^{11}$Li$+p$
elastic scattering}

\begin{figure}
\includegraphics[width=0.8\linewidth]{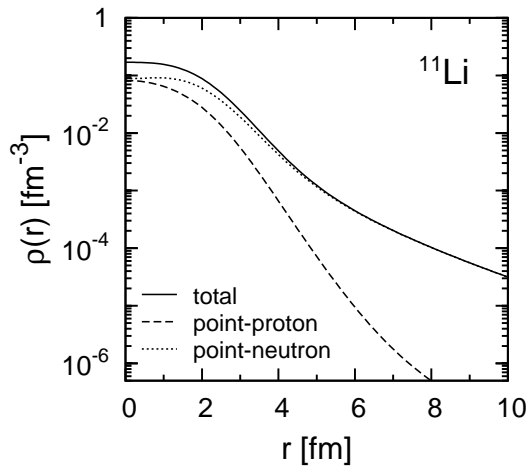}
\caption{Total (normalized to $A=11$), point-proton (normalized to
$Z=3$) and point-neutron (normalized to $N=8$) densities of
$^{11}$Li obtained in the LSSM approach
\protect\cite{Karataglidis97}.
\label{fig1}}
\end{figure}

In the beginning of this subsection we consider $^{11}$Li$+p$ elastic scattering at three energies, 62, 68.4, and 75 MeV/nucleon, for which the differential cross
sections have been measured \cite{Moon92,Korsh97c,Korsh96}. The respective folding optical potentials $V^{F}$ and $W^{H}$ are calculated by the procedure described
in the previous subsections IIA.B.C. using Eqs.~(\ref{eq:0}--\ref{eq:15}) and then, the whole OP is constructed in the form
\begin{widetext}
\begin{equation}
U_{opt}(r)=N_R V^F(r) + i N_I W(r) + 2\lambda_{\pi}^{2} \left\{ N_R^{SO} V^F_0 \frac 1 r \frac {df_R(r)} {dr} + i N_I^{SO} W^H_0 \frac 1 r \frac {df_I(r)}
{dr}\right \} ({\bf l.s}). \label{eq:16}
\end{equation}
\end{widetext}
The OP $U_{opt}(r)$ (\ref{eq:16}) is applied to calculate the
elastic scattering differential cross sections using the program
DWUCK4 \cite{DWUCK}. The number of partial waves is controlled by
the parameter LMAX that corresponds to the maximum partial wave
for the distorted waves. We use the parameter LMAX=100. For the
densities of protons and neutrons of $^{11}$Li we use the LSSM
ones \cite{Karataglidis97} (shown in Fig.~\ref{fig1}) that have an
exponential asymptotics which is the correct one. As can be seen
from Eq.~(\ref{eq:16}), we introduce and consider the set of $N$
coefficients as parameters that can be found by fitting the
calculated to the experimental differential cross sections of the
$^{11}$Li$+p$ elastic scattering. Moreover, the fitting procedure
can be constrained by additional conditions on the behavior of the
OP's (as in Refs.~\cite{Lukyanov2007,Lukyanov2009,Lukyanov2010}
and will be seen below). The real and imaginary parts of the SO
optical potential in (\ref{eq:16}) are approximated by Woods-Saxon
form. Their parameters $V_{0}^{F}$($W_{0}^{H}$), $R_{R}(R_{I})$
and $a_{R}(a_{I})$ were obtained by a fitting procedure to the
respective calculated microscopic potentials $V^{F}(r)$ and
$W^{H}(r)$. We take the ImOP in two forms, the microscopically
obtained $W^{H}$ within HEA ($W=W^{H}$) or the form of the folded
real potential $V^{F}$ ($W=V^{F}$).

Concerning our approach using the set of $N$ coefficients as parameters we consider it as the appropriate physical basis, which constrains the fitting procedure by
the established model forms of the potentials. We emphasize that in our work we do not aim to find perfect agreement with the experimental data. In this sense,
however, the usage of the fitting parameters ($N$'s) related to the depths of the different components of the OP's can be considered as a way to introduce a
quantitative measure of the deviations of the predictions of our method (with the account for the exchange contributions to OP) from the reality (e.g., the
differences of $N$'s from unity for given energies, as can be seen below). Thus, the closeness of the $N$'s values to unity could show the ability of the approach
to give the absolute values of the intensity of the OP's.

\begin{figure}
\includegraphics[width=0.8\linewidth]{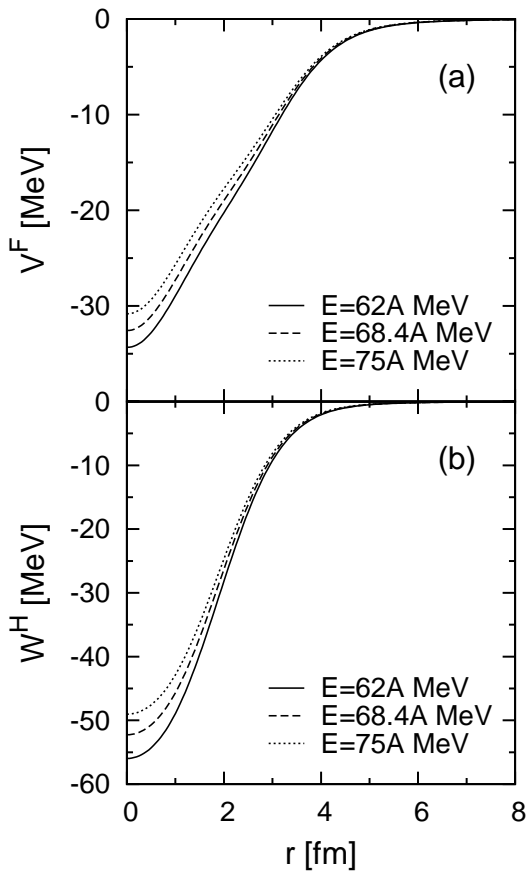}
\caption{Microscopic real part ($V^F$) of OP (a) and HEA imaginary part ($W^H$) (b) calculated using the LSSM densities for energies $E=62$ (solid lines), 68.4
(dashed lines) and 75 MeV/nucleon (dotted lines).
\label{fig2}}
\end{figure}

\begin{figure}
\includegraphics[width=1.0\linewidth]{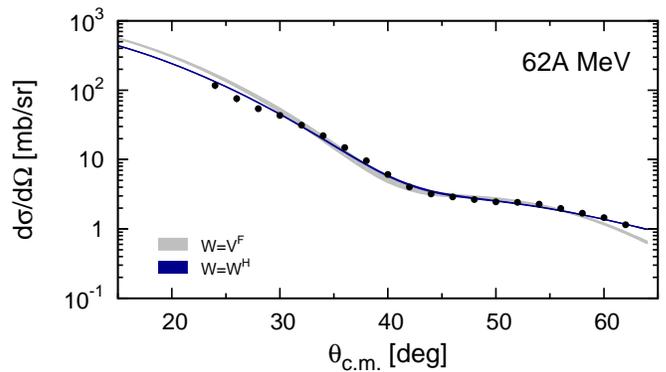}
\caption{(Color online) The $^{11}$Li$+p$ elastic scattering cross section at $E=62$ MeV/nucleon using $U_{opt}$ [Eq.~(\ref{eq:16})] for values of the parameters
shown in Table~\ref{tab1}. Dark (blue) area: $W=W^{H}$, pale (grey) area: $W=V^{F}$. The experimental data are taken from Ref.~\protect\cite{Moon92}.
\label{fig3}}
\end{figure}

The microscopic real part ($V^{F}$) of OP and HEA imaginary part ($W^{H}$) calculated using LSSM densities of $^{11}$Li are shown in Fig.~\ref{fig2} for different
energies. In Fig.~\ref{fig3} we give as an example the differential cross section of the elastic scattering $^{11}$Li$+p$ at 62 MeV/nucleon in the cases when
$W=W^{H}$ and $W=V^{F}$ with and without accounting for the spin-orbit term in Eq.~(\ref{eq:16}). The renormalization parameters $N$ are determined by a fitting
procedure. The results of the calculations are close to each other and that is why all of them are presented inside areas shown in Fig.~\ref{fig3}. The following
definition of $\chi^2$ is used:
\begin{equation}
\chi^2 = \frac{1}{N} \sum\limits_{i=1}^{N} \Bigl[ \frac{\sigma^{\text{exp}}(\vartheta_i) - \sigma^{\text{th}} (\vartheta_i)}{\Delta
\sigma^{\text{exp}}(\vartheta_i)} \Bigr]^2, \label{eq:16a}
\end{equation}
where $\sigma^{\text{th}}(\vartheta_i)$ and $\sigma^{\text{exp}}(\vartheta_i)$ are the theoretical and experimental values of the differential cross sections
($d\sigma/d\Omega$), and $\Delta\sigma^{\text{exp}}(\vartheta_i)$ is the experimental error. The blue area in Fig.~\ref{fig3} includes four curves corresponding to
$W=W^{H}$ (from which three curves obtained without SO term and one with the SO term), while the grey one includes four curves corresponding to $W=V^{F}$ (from
which two curves obtained without SO term and two curves with the SO term). We give in Table~\ref{tab1} the values of the $N$'s parameters, $\chi^{2}$ and the total
reaction cross sections $\sigma_{R}$.

\begin{table}
\caption{Values of the $N$'s parameters, $\chi^{2}$ and $\sigma_{R}$ (in mb) in the case of $^{11}$Li$+p$ at 62 MeV/nucleon for the results shown in
Fig.~\ref{fig3}.} \label{tab1}
\begin{center}
\begin{tabular}{ccccccc}
\hline \hline \noalign{\smallskip}
$W$ & $N_R$ & $N_I$ & $N_R^{SO}$ & $N_I^{SO}$ & $\chi^{2}$ & $\sigma_{R}$ \\
\noalign{\smallskip}\hline\noalign{\smallskip}
$W^{H}$ & 0.871 & 0.953 &       &       & 1.415 & 456.97 \\
        & 0.870 & 0.965 &       &       & 1.435 & 459.37 \\
        & 0.873 & 0.948 &       &       & 1.423 & 455.98 \\
        & 0.854 & 0.974 & 0.028 & 0.000 & 1.468 & 461.21 \\
\noalign{\smallskip}
$V^{F}$ & 0.953 & 0.448 &       &       & 5.567 & 389.72  \\
        & 0.956 & 0.398 &       &       & 5.726 & 361.02  \\
        & 0.670 & 0.251 & 0.338 & 0.000 & 5.027 & 258.65  \\
        & 0.623 & 0.266 & 0.402 & 0.000 & 5.538 & 270.05  \\
\noalign{\smallskip}\hline \hline
\end{tabular}
\end{center}
\end{table}

It can be seen from Fig.~3 the satisfactory overall agreement of both areas of curves with the experimental data. However, we note the better agreement in the case
when $W=W^{H}$ (the blue area) and the values of $\chi^{2}$ are between 1.40 and 1.47, while in the case $W=V^{F}$ they are between 5.00 and 5.80. The situation is
similar also for the other energies. So, in our further calculations we use only ImOP $W=W^{H}$. Second, we note that the values of $\sigma_{R}$ are quite different
in both cases ($\sigma_{R}\approx $ 455--462 mb for $W=W^{H}$ and $\sigma_{R}\approx $ 260--390 mb for $W=V^{F}$). Third, one can see from Table~\ref{tab1} and from
the comparison with the data in Fig.~\ref{fig3} that the role of the SO term is weak. Its effects turn out to be to decrease the values of $N_{R}$ and to increase
the values of $N_{R}^{SO}$ (see the last two lines in Table~\ref{tab1}).

As is known, the problem of the ambiguity of the parameters $N$ arises when the fitting procedure is applied to a limited number of experimental data (see, e.g.,
the calculations and discussion in our previous works \cite{Lukyanov2007,Lukyanov2009,Lukyanov2010}). Due to the fact that the fitting procedure belongs to the
class of the ill-posed problems (see, e.g., Ref.~\cite{Tikhonov77}), it becomes necessary to impose some physical constraints on the choice of the set of parameters
$N$. The total cross section of scattering and reaction is one of them, however, the corresponding experimental values are missing at the energy interval considered
in the present work.

Another physical criterion that has to be imposed on the choice of the $N$ values is the behavior of the volume integrals
\begin{equation}
J_V=\frac{4\pi}{A}\int dr r^2 [N_{R}V^{F}(r)], \label{eq:17}
\end{equation}
\begin{equation}
J_W=\frac{4\pi}{A}\int dr r^2 [N_{I}W^{H}(r)] \label{eq:18}
\end{equation}
as functions of the energy.

\begin{figure}
\includegraphics[width=1.0\linewidth]{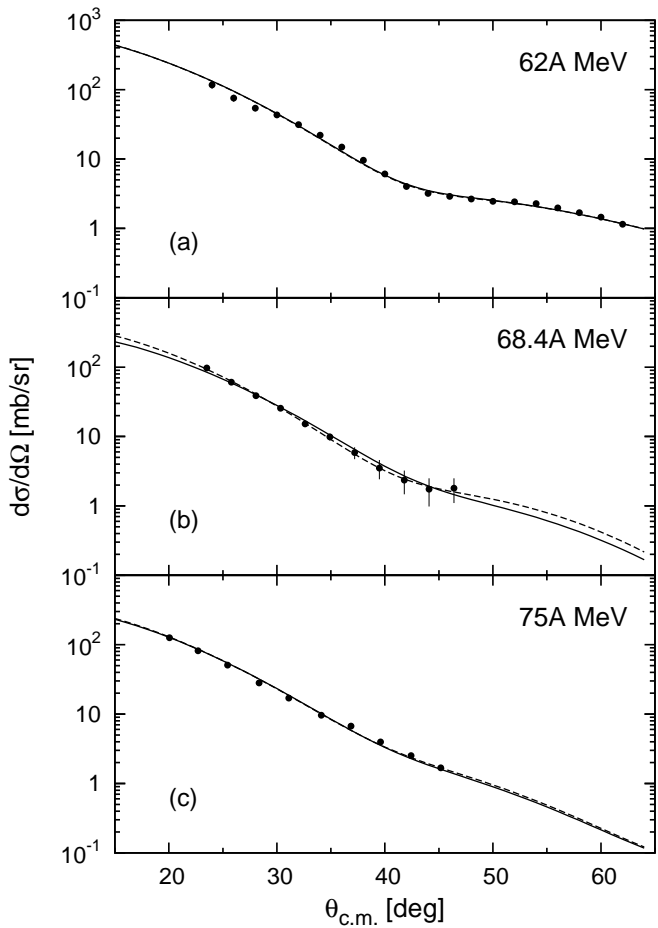}
\caption{The $^{11}$Li$+p$ elastic scattering cross section at $E=62$, 68.4, and 75 MeV/nucleon. Solid line: without SO term; dashed line: with SO term. The values
of $N$'s are given in Table~\ref{tab2}. The experimental data are taken from \protect\cite{Moon92} for 62 MeV/nucleon, \protect\cite{Korsh97c} for 68.4 MeV/nucleon,
and \protect\cite{Korsh96} for 75 MeV/nucleon.
\label{fig4}}
\end{figure}

\begin{figure}
\includegraphics[width=0.8\linewidth]{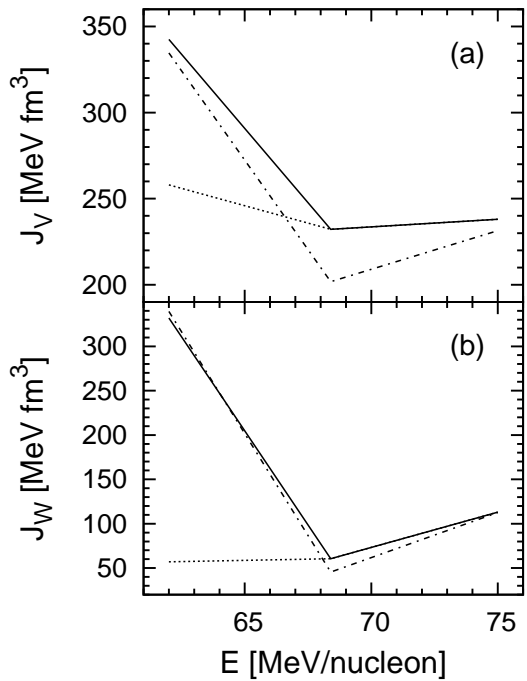}
\caption{The values of the volume integrals $J_V$ and $J_W$ [Eqs.~(\ref{eq:17}) and (\ref{eq:18})] as functions of the energy per nucleon for $^{11}$Li$+p$ elastic
scattering. The $N$'s values are given in Table~\ref{tab2}. Solid line: without SO term of $U_{opt}$ [Eq.~(\ref{eq:16})]; dash-dotted line: with SO term of
$U_{opt}$. The additional values of $J_V$ and $J_W$ at $E=62$ MeV/nucleon (connected by a dotted line with the other curves) are obtained in the case when the
fitting procedure for the $N$'s parameters is limited up to the experimental points for $\theta_{c.m.}\leq 46^\circ$ (see the text).
\label{fig5}}
\end{figure}

\begin{figure}
\includegraphics[width=0.8\linewidth]{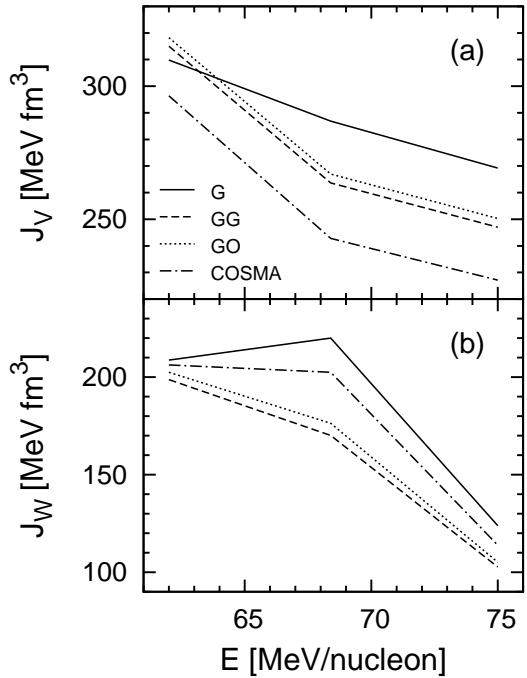}
\caption{The energy dependence of the volume integrals: (a) $J_V$ obtained in \protect\cite{Hassan2009} for folding potentials ReOP ($V$) calculated using two types
of Gaussians (G and GG), Gaussian oscillator (GO) and COSMA densities of $^{11}$Li for $^{11}$Li+$p$ elastic scattering; (b) $J_W$ calculated using the fitted
imaginary WS potentials corresponding to those real parts of OP that give $J_{V}$'s in (a).
\label{fig6}}
\end{figure}

We show in Fig.~\ref{fig4} the results of our calculations of the $^{11}$Li$+p$ elastic scattering cross sections for the three energies $E=62, 68.4$ and 75
MeV/nucleon. For each energy we present two curves, with and without accounting for the SO term. The corresponding values of the $N$'s parameters together with
those of $J_V$, $J_W$, $\chi^{2}$ and $\sigma_{R}$ are given in Table~\ref{tab2}. In Fig.~\ref{fig5} we give the curves for the volume integrals $J_V$ and $J_W$
connecting the results obtained in our calculations with $N$'s values. We present them as better ones because first, the values of $\chi^{2}$ are around unity, and
second, there is a good agreement with the data including those of $\theta_{c.m.}$ up to 60$^{\circ}$ for 62 MeV/nucleon. One can see that the values of $J_V$ are
decreasing with the increase of the incident energy (with a small exception at 68.4 MeV/nucleon) that is in general agreement with the results from
Ref.~\cite{Romanovsky}. This is not the case for $J_W$, where its value for $E=62$ MeV/nucleon is larger than for the others. Indeed, it was pointed out in
\cite{Romanovsky} that the general behavior of the volume integral $J_V$ is decreasing with the increase of the energy in the interval $0<E<100$ MeV/nucleon, while
$J_W$ increases with the increase of comparatively small energy and becomes almost constant at a larger energy. However, the same situation had appeared in the
analysis of the same data at three energies within the semi-microscopic approach in Ref.~\cite{Hassan2009}, where the ReOP was calculated using a single-folding
procedure with Gaussian, Gaussian-oscillator and COSMA forms of the single-particle density and the ImOP was taken phenomenologically in a Woods-Saxon form or equal
to the form of the folded ReOP. In Fig.~\ref{fig6}(a) are shown the curves of $J_V$ corresponding to its values obtained in \cite{Hassan2009} for the cases of the
four densities used. In addition, we show in Fig.~\ref{fig6}(b) the values of $J_W$ calculated using the corresponding fitted imaginary part of the OP's taken in a
phenomenological WS form. One can see that the $J_V$ has a reasonable behavior in agreement with the results from Ref.~\cite{Romanovsky}, while the values of $J_W$
are in contradiction with them. Thus, the problem arising in our work had appeared also in the semi-phenomenological approach in Ref.~\cite{Hassan2009}, in which a
larger number of parameters has been used. A possible reason for such a behavior of $J_W$ at this energy could be the change of the scattering mechanism with the
increase of the angle of scattering when the other channels except the elastic one should be taken into consideration. Such a "strong" channel with its influence on
the elastic one could be that of the fragmentation of $^{11}$Li into clusters.

\begin{table*}
\caption{Values of the $N$'s parameters, volume integrals $J_V$ and $J_W$ (in MeV fm$^{3}$), $\chi^{2}$ and total reaction cross section $\sigma_{R}$ (in mb) for
results at three energies $E$ (in MeV/nucleon) considered and shown in Fig.~\ref{fig4}.} \label{tab2}
\begin{center}
\begin{tabular}{lcccccrcc}
\hline \hline \noalign{\smallskip}
$E$ & $N_R$ & $N_I$ & $N_R^{SO}$ & $N_I^{SO}$ & $J_V$ & $J_W$ & $\chi^{2}$ & $\sigma_{R}$ \\
\noalign{\smallskip}\hline\noalign{\smallskip}
62   & 0.871 & 0.953 &       &       & 342.474 & 332.015 & 1.415 & 456.97 \\
     & 0.851 & 0.974 & 0.028 & 0.000 & 334.610 & 339.332 & 1.468 & 461.21 \\
\noalign{\smallskip}
68.4 & 0.625 & 0.186 &       &       & 232.210 &  60.489 & 1.328 & 153.44 \\
     & 0.543 & 0.140 & 0.201 & 0.000 & 201.744 &  45.530 & 0.316 & 122.25 \\
\noalign{\smallskip}
75   & 0.679 & 0.370 &       &       & 238.048 & 112.913 &       & 232.62 \\
     & 0.660 & 0.369 & 0.045 & 0.000 & 231.387 & 112.607 &       & 232.62 \\
\noalign{\smallskip}\hline \hline
\end{tabular}
\end{center}
\end{table*}

\begin{figure}
\includegraphics[width=1.0\linewidth]{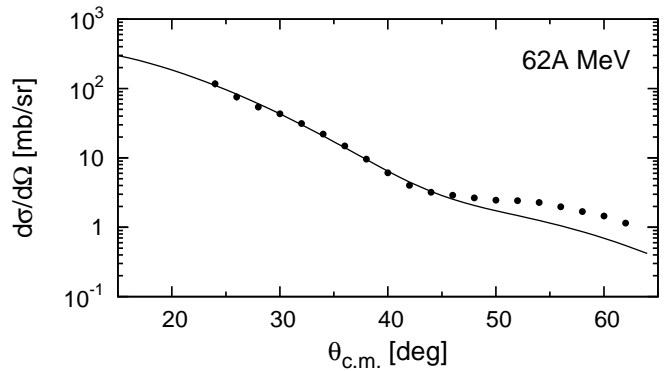}
\caption{The $^{11}$Li$+p$ elastic scattering cross section at $E=62$ MeV/nucleon when the fitting procedure for the $N$'s parameters is limited only up to the
experimental points for $\theta_{c.m.}\leq 46^\circ$. The obtained values of $N_R$, $N_I$, $J_V$, $J_W$, $\chi^{2}$, and $\sigma_{R}$ are given in the text.
\label{fig7}}
\end{figure}

As a next step, we perform a methodical study of $^{11}$Li$+p$ elastic scattering cross section for $E=62$ MeV/nucleon limiting our fitting procedure for the $N$'s
parameters up to the experimental points for $\theta_{c.m.}\leq 46^\circ$. The result of this study is presented in Fig.~\ref{fig7}. Doing so we consider now the
experimental data for all three energies 62, 68.4, and 75 MeV/nucleon being at the same region of angles. The fit to this amount of data at 62 MeV/nucleon yields
the new set of parameters: $N_R=0.656$, $N_I=0.164$ with $\chi^{2}=0.788$ and $\sigma_{R}=154.86$ mb. Now we obtain values of the volume integrals (without SO term
of $U_{opt}$) $J_V=257.973$ MeV fm$^{3}$ and $J_W=57.136$ MeV fm$^{3}$ (shown in Fig.~\ref{fig5}), while the obtained before values are $J_V=342.47$ MeV fm$^{3}$
and $J_W=332.015$ MeV fm$^{3}$ (see the first line in Table~\ref{tab2}). As a result, we get the behavior of $J_V$ and $J_W$ in a reasonable agreement with the
conclusions of Ref.~\cite{Romanovsky}. In our opinion, the procedure described above points out the role of the data at $\theta_{c.m.}>46^\circ$ on the values of
$\chi^{2}$ and on the conclusions on the mechanism of the elastic scattering process.

\section{Breakup processes within $^{9}$Li+$2n$ cluster model}

\subsection{Two-cluster model and applications}

In this Section, in addition to the analysis of $^{11}$Li$+p$ elastic scattering cross section in Sec.~II, we study other characteristics of the reaction mechanism,
such as the $^{11}$Li total reaction cross section, the breakup cross section and related quantities. This part of the work is based on the procedure for
microscopic calculations of OP's presented in Sec.~II. We consider a simple two-cluster model that has been already used for $^{6}$He for studying its elastic
scattering and breakup reactions on nuclear targets \cite{Lukyanov2011}. Within this model for the $^{11}$Li nucleus, first, the density distributions of $^{9}$Li
core ($c$-cluster) and $h=2n$-halo must be given. Second, the folding potentials of interaction of each of the clusters with the incident proton have to be
computed. Then, the sum of these two potentials must be folded with the respective two-cluster density distribution of $^{11}$Li that causes the necessity the wave
function of the relative motion of two clusters to be known. We calculate the latter as a solution of the Schr\"{o}dinger wave equation by using of WS potential and
given 0$s$ or 1$s$ state for particle with reduced mass of both clusters. The WS parameters are obtained by fitting the energy of a given state to the empirical
separation energy value of $h$-cluster $\varepsilon=0.247$ MeV and the rms radius of the cluster function. For the latter we choose the value of 4.93 fm that is
somehow "in between" the values obtained within the three-body COSMA \cite{Thompson94} and deduced from shell-model calculations \cite{Zelenskaya,Myo2007}. Such
two-cluster model takes an interspace between the two classes of approaches. In one of them each of the clusters has its own phenomenological density that is often
used to fit the elastic scattering data. The second class includes microscopic three-body models using to a different extent the shell-model picture. Among them we
would like to note COSMA (see, for example, Refs.~\cite{Tostevin98,Ershov2010}), which has already successfully described a great amount of experimental data
applying the Glauber scattering theory. Justifying our more simple two-cluster model, we hope, however, to keep the basic physical consideration avoiding some
simplifications like folding without exchange effects, use of Gaussian-type functions for densities of clusters and bound-state wave functions of relative motion,
use of phenomenological ImOP etc. We will always take into account the contribution of the exchange effects and the wave function of the relative motion of two
clusters is calculated for the fitted finite-range potential that has an exponential behavior. The bound-state two-cluster system requires a particular
consideration. In the earlier works estimations were made using the wave function of 0$s$-state ($n$=0) of the $(c+h)$ system, which does not have nodes inside the
potential (except at $r=0$). However, it has been shown in Refs.~\cite{Thompson94,Myo2007} that due to the violation of Pauli principle (Pauli-blocking effect in
$^{11}$Li ground state) the 1$s$- and 0$p$-states give the main contribution to the wave function of the two-cluster system with almost equal probabilities thus
oscillating once inside the potential. Nevertheless, we will consider both $0s$- and $1s$-densities $\rho_{0}^{(0)}$ and $\rho_{0}^{(1)}$ in the further
calculations and comparisons of the results.

In the present study, the interaction between the clusters is
taken to be a WS potential with the adjusted geometrical
parameters $R=1.0$ fm, $a=0.25$ fm and the depth $V_{0}=32.55$ MeV
for 0$s$-state and $R=6.25$ fm, $a=0.25$ fm, and $V_{0}=11.55$ MeV
for 1$s$-state.

The $s$-state ($l=0$) wave function of the relative motion of two clusters is
\begin{equation}
\phi_{00}^{(n)}({\bf s})=\phi_{0}^{(n)}(s)\frac{1}{\sqrt{4\pi}}, \;\;\; n=0, 1
\label{eq:15a}
\end{equation}
and thus, the respective density distribution is defined as a probability for clusters to be at a mutual distance $s$:
\begin{equation}
\rho_{0}^{(n)}({\bf s})=|\phi_{00}^{(n)}({\bf s})|^{2}= \frac{1}{4\pi}|\phi_{0}^{(n)}(s)|^{2}.
\label{eq:15c}
\end{equation}

In the framework of the $^{9}$Li+2$n$ model of $^{11}$Li one can
estimate the $^{11}$Li$+p$ OP as a sum of two OP's of interactions
of the $c$- and $h$-clusters with protons folded with the density
$\rho_{0}^{(n)}(s)$ ($n$=0, 1):
\begin{widetext}
\begin{eqnarray}
U^{(b,n)}(r)&=&V^{(b,n)}+iW^{(b,n)}=\int d{\bf
s}\rho_{0}^{(n)}(s)\left [U_{c}^{(n)}\left ({\bf r}+(2/11){\bf
s}\right )+U_{h}^{(n)}\left ({\bf r}-(9/11){\bf
s}\right )\right ]=2\pi \int_{0}^{\infty} \rho_{0}^{(n)}(s)s^{2}ds \nonumber \\
&\times & \int_{-1}^{1} dx \left
[U_{c}^{(n)}\left(\sqrt{r^{2}+(2s/11)^{2}+r(4/11)sx}\right )+
U_{h}^{(n)}\left(\sqrt{r^{2}+(9s/11)^{2}-r(18/11)sx}\right )\right
]. \label{eq:15d}
\end{eqnarray}
\end{widetext}
In Eq.~(\ref{eq:15d}) ${\bf r}-(9/11){\bf s}\equiv {\bf r}_{h}$
and ${\bf r}+(2/11){\bf s}\equiv {\bf r}_{c}$ define the
corresponding distances between the centers of each of the
clusters and the arbitrary position of the nucleon in $^{11}$Li
nucleus, and ${\bf s}={\bf s}_{1}+{\bf s}_{2}=(9/11){\bf
s}+(2/11){\bf s}$ determines the relative distance between the
centers of the two clusters, $s_{1}$ and $s_{2}$ being distances
between the centers of $^{11}$Li and each of the clusters,
respectively. The potential $U_{c}^{(n)}$ in Eq.~(\ref{eq:15d}) is
calculated within the microscopic hybrid model of OP described in
Sect.IIA and B. For OP of the $h$-$p$ interaction we use the sum
of two $v_{np}$ potentials as
\begin{equation}
U_{h}^{(n)}=2v_{np}=2v(r)(1+i\gamma).
\label{eq:15e}
\end{equation}
Such $n$-$p$ complex potential has been used in the four-body
model \cite{Suzuki93} in calculations of the $^{11}$Li$+p$ elastic
scattering and it was shown that the cross sections are rather
insensitive to a precise form of the $n$-$p$ potential taken in
the form \cite{Thompson77} (in MeV):
\begin{equation}
v(r)=120e^{-1.487r^{2}}-53.4e^{-0.639r^{2}}-27.55e^{-0.465r^{2}}
\label{eq:15f}
\end{equation}
with $\gamma=0.4$.

We also intend to adopt the two-cluster model to calculate breakup
reactions of $^{11}$Li in collisions with the proton target. To
this end the HEA method which has been developed in
Refs.~\cite{Hencken96,Bertulani2004} and applied in
\cite{Lukyanov2011} for $^{6}$He+$^{12}$C reaction will be used in
the present study. For simplicity, further the superscript index
($n$=0, 1) which corresponds to the number of nodes of the
relative-motion $s$-wave function of the two clusters will be
omitted. To show briefly the eikonal formalism, we start with the
probability that after the collision with a proton ($z\rightarrow
\infty $) the cluster $h$ or $c$ with an impact parameter $b$
remains in the elastic channel:
\begin{widetext}
\begin{eqnarray}
|S_{i}(b)|^{2}=\exp{\left[-\frac{2}{\hbar
v}\int_{-\infty}^{\infty}dzW_{i}\left(\sqrt{b^{2}+z^{2}}\right
)\right ]},\;\;\;\;\; i=c,h,
\label{eq:15g}
\end{eqnarray}
\end{widetext}
where $W$ is the imaginary part of the microscopic OP
(\ref{eq:15d}). Consequently, the probability for the cluster to
be removed from the elastic channel is $(1-|S|^{2})$. Thus, the
common probability of both $h$ and $c$ clusters to leave the
elastic channel of the $^{11}$Li$+p$ scattering is
$(1-|S_{h}|^{2})(1-|S_{c}|^{2})$. After averaging the latter by
$\rho_{0}(s)$ (which characterizes the probability of $h$ and $c$
to be at a relative distance $s$), the total absorbtion cross
section is obtained:
\begin{equation}
\sigma_{abs}^{tot}=2\pi \int_{0}^{\infty}b_{h}db_{h}
[1-|S_{h}(b_{h})|^{2}] [1-I_{c}(b_{h})],
\label{eq:15h}
\end{equation}
where
\begin{equation}
I_{c}(b_{h})=\int d{\bf s}\rho_{0}(s)|S_{c}(b_{c})|^{2}.
\label{eq:15i}
\end{equation}
In Eq.~(\ref{eq:15i})
\begin{equation}
b_{c}=\sqrt{s^{2}\sin^{2}\theta+b_{h}^{2}+2sb_{h}\sin\theta\cos(\varphi-
\varphi_{h})}
\label{eq:15ia}
\end{equation}
and it comes out from the relation ${\bf b}_{c}={\bf b}_{h}+{\bf
b}$ with $b=s\sin\theta$ being the projection of ${\bf s}$ on the
plane normal to the $z$-axis along the straight line trajectory of
the incident nucleus.

In the case of a stripping reaction with removing $h$-cluster from
$^{11}$Li to the proton target, one should use the probability of
$h$ to leave the elastic channel $[1-|S_{h}(b_{h})|^{2}]$, and for
$c$ to continue its elastic scattering with a probability
$|S_{c}(b_{c})|^{2}$. Then the probability of the whole process is
$|S_{c}(b_{c})|^{2}[1-|S_{h}(b_{h})|^{2}]$, and to get the total
stripping cross section one has to average over $\rho_{0}(s)$ [see
Eqs.~(\ref{eq:15h}) and (\ref{eq:15i})]. Similarly, the $^{9}$Li
transfer can be constructed, and the net contribution of both
removal reactions yields the total breakup cross section:
\begin{eqnarray}
\sigma_{bu}^{tot} &=& 2\pi \int_{0}^{\infty}
b_{h}db_{h}\{|S_{h}(b_{h})|^{2} \nonumber \\
&+& [1-2|S_{h}(b_{h})|^{2}]I_{c}(b_{h})\}.
\label{eq:15j}
\end{eqnarray}
The sum of both absorption [Eqs.~(\ref{eq:15h}) and
(\ref{eq:15i})] and breakup [Eq.~(\ref{eq:15j})] cross sections
gives the total reaction cross section:
\begin{equation}
\sigma_{R}^{tot}=2\pi \int_{0}^{\infty}
b_{h}db_{h}[1-|S_{h}(b_{h})|^{2}I_{c}(b_{h})].
\label{eq:15k}
\end{equation}

\subsection{Momentum distributions of fragments}

As is known (see, e.g., \cite{Hencken96}), the differential and
total cross sections (for elastic scattering, as well as for
diffractive breakup and absorption) all require calculations of
the probability functions of the ${\bf k}$-momentum distribution
of a cluster in the two-cluster system $d^{3}P({\bf b},{\bf
k})/d{\bf k}$ that depend on the impact parameter $\bf b$. The
general expression for the probability functions can be written as
\cite{Hencken96}:
\begin{equation}
\frac{d^{3}P_{\Omega}({\bf b},{\bf k})}{d{\bf k}}=\frac{1}{(2\pi)^{3}}
\left |\int d{\bf s} \phi_{{\bf k}}^{*}({\bf s})\Omega({\bf b},{\bf r}_{\perp})
\phi_{00}^{(n)}({\bf s})\right |^{2},
\label{eq:15m}
\end{equation}
where  $\Omega({\bf b},{\bf r}_{\perp})$  is expressed by means of
the two profile functions $S_{c}$ and $S_{h}$ [Eq.~(\ref{eq:15g})]
of the core and the di-neutron clusters, respectively. In
Eq.~(\ref{eq:15m}) $\phi_{{\bf k}}({\bf s})$ is the continuum wave
function and ${\bf k}$ is the relative momentum of both clusters
in their center-of-mass frame. The vector ${\bf r}_{\perp}$ is the
projection of the relative coordinate ${\bf s}$ between the
centers of the two clusters on the plane normal to the $z$-axis
mentioned above. The ground-state wave function of the relative
motion of the two clusters $\phi_{00}$ is given for the $s$-state
by Eq.~(\ref{eq:15a}). For calculations of e.g., the diffractive
cross sections, the continuum wave function $\phi_{\bf k}$ is
expanded in partial wave representation. If in this case the
distortion in the final channel is neglected, the wave function
$\phi_{{\bf k}}({\bf s})$ is replaced by a plane wave. Then,
following Ref.~\cite{Hencken96} for the $s$-state ($l=0$) the
expression for $d^{2}P_{\Omega}({\bf b},{\bf k})/dk_{L}dk_{\perp}$
will take the form:
\begin{widetext}
\begin{equation}
\frac{d^{2}P_{\Omega}({\bf b},{\bf k})}{dk_{L}dk_{\perp}}=
\frac{k_{\perp}}{16\pi^{3}k^{2}}\left |\int ds \int
d(\cos\theta_{s})\,g(s)\sin{(ks)}\int d\varphi_{s}\Omega({\bf
b},{\bf r}_{\perp})\right |^{2}
\label{eq:15s}
\end{equation}
\end{widetext}
with
\begin{equation}
\Omega({\bf b},{\bf r}_{\perp})=S_{c}({\bf b}_{c})S_{h}({\bf b}_{h}).
\label{eq:15t}
\end{equation}
In Eq.~(\ref{eq:15s})
$g(s)=r\phi_{0}^{(n)}(s)=r\sqrt{4\pi\rho_{0}^{(n)}(s)}$, where
$\phi_{0}^{(n)}$ and $\rho_{0}^{(n)}$ are given in
Eqs.~(\ref{eq:15a}) and (\ref{eq:15c}). Hence, the diffraction
breakup cross section has the form
\begin{widetext}
\begin{equation}
\left (\frac{d\sigma}{dk_{L}}\right )_{diff}=\int_{0}^{\infty} b_h
db_{h}\int_{0}^{2\pi} d \varphi_{h}\int_{0}^{\infty} d{k}_{\perp}
\frac{d^{2}P_{\Omega}({\bf k},{\bf b})}{dk_{L} dk_{\perp}}
\label{eq:15ldiff}
\end{equation}
\end{widetext}
with $d^{2}P_{\Omega}({\bf b},{\bf k})/dk_{L}dk_{\perp}$ from
Eq.~(\ref{eq:15s}). The integrations over $b_h$ and $\varphi_{h}$
in Eq.~(\ref{eq:15ldiff}) mean integration over the impact
parameter ${\bf b}_h$ of the cluster $h$ with respect to the
target.

In the case of  the stripping reaction  when the $h$ - cluster
leaves the elastic channel  it can be shown (following
\cite{Hencken96}) that the cross section takes the form:
\begin{widetext}
\begin{equation}
\left(\frac{d\sigma}{dk_{L}}\right)_{str}=\frac{1}{2\pi^{2}}\int_{0}^{\infty}b_{h}db_{h}d\varphi_{h}
\left [ 1-|S_{h}(b_{h})|^{2}\right ] \int \rho d\rho
d\varphi_{\rho} |S_{c}(b_{c})|^{2} \left [ \int_{0}^{\infty}dz
\cos (k_{L}z)\phi_{0}\left (\sqrt{\rho^{2}+z^{2}}\right ) \right
]^{2}.
\label{eq:str}
\end{equation}
\end{widetext}
Eq.~(\ref{eq:str}) is obtained when the incident nucleus has spin
equal to zero and for the $s$-state of the relative motion of both
clusters in the nucleus expressed by Eq.~(\ref{eq:15a}) with ${\bf
s} ={\bf r}_{c}- {\bf r}_{h}$, ${\bf \rho}={\bf b}_{c}-{\bf
b}_{h}$, ${\bf s}={\bf \rho} + {\bf z}$ and $b_{c}$ from
Eq.~(\ref{eq:15ia}).

\subsection{Results of calculations for breakup processes}

\begin{figure}
\includegraphics[width=0.8\linewidth]{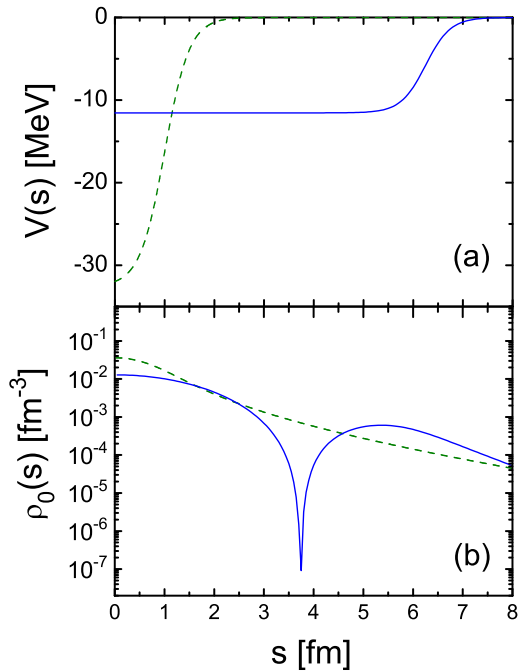}
\caption{(Color online) The WS potential $V(s)$ of the interaction
between $c$ and $h$ clusters (a) and the two-cluster density
distribution $\rho_{0}(s)$ normalized to unity (b) for the cases
of $n$=0 (green dashed line) and $n$=1 (blue solid line).
\label{fig8}}
\end{figure}

To estimate the $^{11}$Li breakup on a proton target, we use the two-cluster model described in Sec.~IIIA. As presented there, we intend to study some observables
when the $^{11}$Li nucleus with the $h=2n$-cluster separation energy of 0.247 MeV is considered as a system in the $l=0$ state with principal quantum numbers $n=0$
or $n=1$. The respective WS potentials $V(s)$ and probabilities $\rho_{0}^{(n)}(s)$ [Eq.~(\ref{eq:15c})] for the distance $s$ between the clusters in $^{11}$Li are
shown for both $n=0, 1$ in Fig.~\ref{fig8}(a) and (b), respectively. It can be seen from Fig.~\ref{fig8}(a) that the WS potential for $n=0$ is about 2.8 times
deeper than the one for the case of $n=1$, although the shapes of both potentials are similar. We note also that the half radius of the $n=1$ potential is equal to
6.25 fm and it is much larger than that of 1.01 fm of the $n=0$ potential. Fig.~\ref{fig8}(b) shows that the two densities differ from each other. Particularly, a
steep drop of $n$=1 density is observed at $s\approx $ 3.8 fm. Moreover, bearing in mind the results of fitting procedures in phenomenological potentials (e.g., in
Ref.~\cite{Dobrovolsky2006}) giving rms radius of about $5\div 6$ fm for the constituent $h$-cluster density $\rho_{h}(r)$, we may conclude that in our
consideration the $n=1$ cluster state of $^{11}$Li becomes preferable. On the other hand, the existence of long tails of $\rho_{0}^{(n=0, 1)}(s)$ for both states
provokes interest to test their effects in the further considerations.

\begin{figure}
\includegraphics[width=1.0\linewidth]{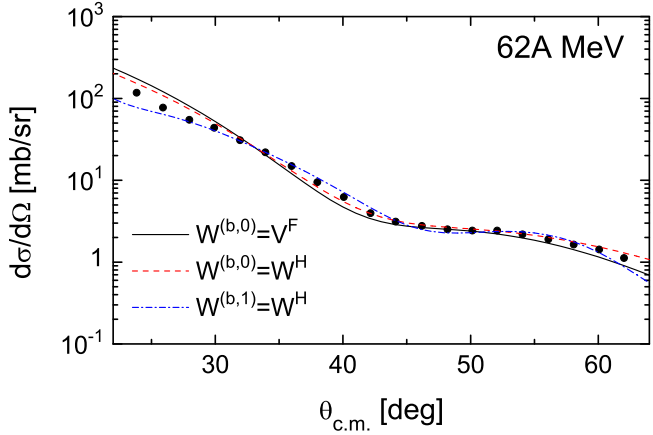}
\caption{(Color online) The $^{11}$Li$+p$ elastic scattering cross
section at $E=62$ MeV/nucleon using $U^{(b,n)}$
[Eq.~(\ref{eq:15d})] for values of the parameters $N$ shown in
Table~\ref{tab3}. Black solid line: $W^{(b,0)}=V^{F}$, red dashed
line: $W^{(b,0)}=W^{H}$, and blue dash-dotted line:
$W^{(b,1)}=W^{H}$. The experimental data are taken from
Ref.~\protect\cite{Moon92}.
\label{fig9}}
\end{figure}

Our next step is to apply the optical potential $U^{(b,n)}$
[Eq.~(\ref{eq:15d})] constructed in the framework of the
two-cluster model of $^{11}$Li to calculate the differential cross
section of the elastic scattering $^{11}$Li$+p$ at 62 MeV/nucleon.
For the real part $V^{(b,n)}$ of this OP we use a single-folding
procedure in which the LSSM density \cite{Karataglidis97} is taken
for the $^{9}$Li cluster. The imaginary part $W^{(b,n)}$ of the OP
is considered like before to be either $W=W^{H}$ or $W=V^{F}$. The
calculated cross sections are shown in Fig.~\ref{fig9} and
compared with the experimental data \cite{Moon92}. For both cases
we give in Table~\ref{tab3} the values of the fitted
renormalization coefficients $N$'s and the respective total cross
sections for $n=0$ and $n=1$ cases. One can see from
Fig.~\ref{fig9} that the angular distributions for both kinds of
ImOP are closely displayed and they lead to a fairly good
agreement with the experimental data. However, we note that the
data are reproduced better again when $W=W^{H}$ for both $n$=0, 1
cases, as it was pointed out from the discussion of the results
presented in Fig.~\ref{fig3} and obtained with the usage of the
LSSM density for $^{11}$Li.

In Table~\ref{tab3} the values of the total absorption
$\sigma_{abs}^{tot}$, breakup $\sigma_{bu}^{tot}$ and total
reaction $\sigma_{R}^{tot}$ cross sections are listed. First, we
note the significant role that the breakup channel plays in the
$^{11}$Li$+p$ reaction, where $\sigma_{bu}^{tot}$ contributes more
than 80\% to $\sigma_{R}^{tot}$. This is not the case of
$^{6}$He+$^{12}$C process at energy of 38.3 MeV/nucleon
\cite{Lukyanov2011}, for which the breakup cross section
constitutes only about the half of the total reaction cross
section. This can be related with the observation that a quite
substantial amount of the $^{11}$Li$+p$ imaginary potential in the
elastic scattering channel is formed due to a transfer of the
incident flux of $^{11}$Li to a larger extent into breakup
channels. Also, for the case of $n=1$ state of the cluster wave
function, the fitted strength coefficients $N$'s and the
respective values of the cross sections are larger than for the
$n=0$ state, but the general conclusions on the preferable role of
breakup processes remain the same.

\begin{table}
\caption{The $N$'s parameters of OP's for $^{11}$Li$+p$ scattering
at 62 MeV/nucleon and HEA estimations of the total cross sections
$\sigma_{abs}^{tot}$ [Eq.~(\ref{eq:15h})], $\sigma_{bu}^{tot}$
[Eq.~(\ref{eq:15j})], and $\sigma_{R}^{tot}$ [Eq.~(\ref{eq:15k})]
(in mb) using the cluster model of $^{11}$Li.} \label{tab3}
\begin{center}
\begin{tabular}{cccccc}
\hline \hline \noalign{\smallskip}
$W^{(b,n)}$ & $N_R$ & $N_I$ & $\sigma_{abs}^{tot}$ & $\sigma_{bu}^{tot}$ & $\sigma_{R}^{tot}$ \\
\noalign{\smallskip}\hline\noalign{\smallskip}
$W^{(b,0)}=V^{F}$  & 1.407 & 1.195 & 79.0  & 431.8 & 510.8 \\
$W^{(b,0)}=W^{H}$  & 1.381 & 1.306 & 78.6  & 405.3 & 483.9 \\
$W^{(b,1)}=W^{H}$  & 4.68  & 3.99  & 106.6 & 581.6 & 688.2 \\
\noalign{\smallskip}\hline \hline
\end{tabular}
\end{center}
\end{table}
\begin{figure}
\includegraphics[width=1.0\linewidth]{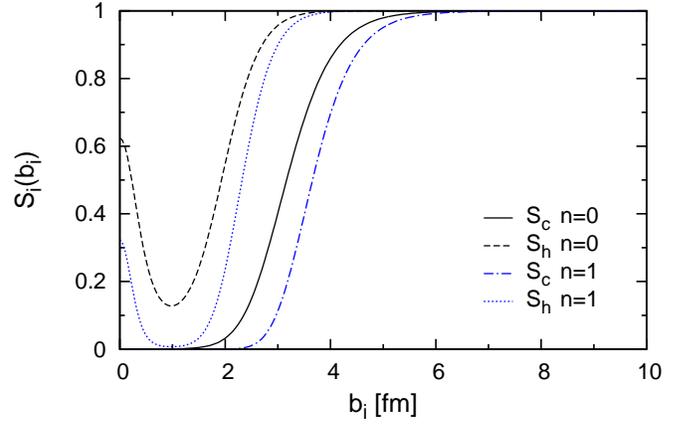}
\caption{(Color online) The functions $S_{i}(b_{i})$, $i=c,h$ [see Eq.~(\ref{eq:15g})] for $s$-state of the relative motion of clusters with $n$=0 and $n$=1.
\label{fig10}}
\end{figure}
\begin{figure}
\includegraphics[width=1.0\linewidth]{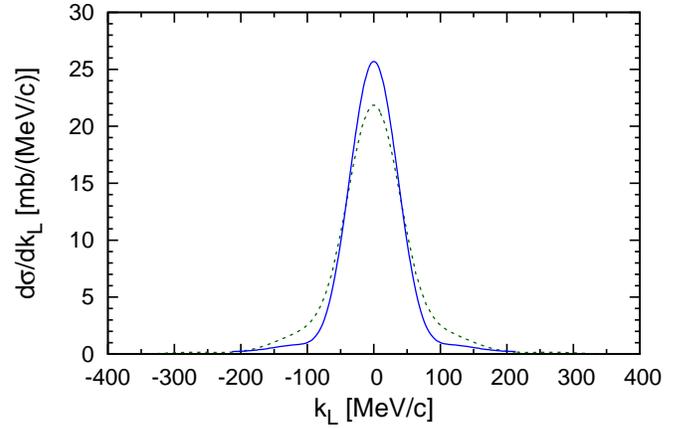}
\caption{(Color online) Cross section of diffraction breakup in $^{11}$Li$+p$ scattering at $E=62$ MeV/nucleon for the cases of $n$=0 (green dashed line) and $n$=1
(blue solid line).
\label{fig11}}
\end{figure}

\begin{figure}
\includegraphics[width=1.0\linewidth]{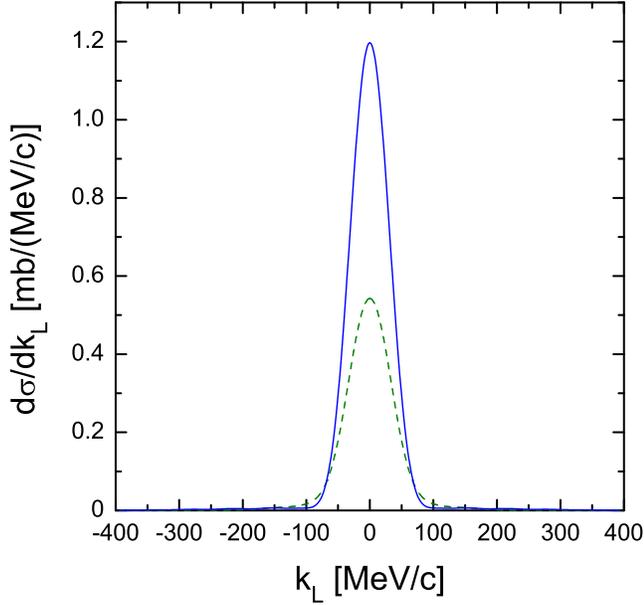}
\caption{(Color online) The same as in Fig.~\ref{fig11} but for the stripping reaction.
\label{fig12}}
\end{figure}

Our next step is to calculate using Eqs.~(\ref{eq:15ldiff}) and (\ref{eq:str}) as examples the cross sections of the diffractive and stripping (when $h=2n$ cluster
leaves the elastic channel) $^{11}$Li$+p$ reactions at $E=62$ MeV/nucleon, respectively. For this purpose we use in Eqs.~(\ref{eq:15ldiff}) and (\ref{eq:str}) the
corresponding functions $S_{i}(b_{i})$, $i=c,h$ [see Eq.~(\ref{eq:15g})]. They are given in Fig.~\ref{fig10} for $s$-state with $n$=0 and $n$=1. In
Figs.~\ref{fig11} and \ref{fig12} we show the results for the diffraction breakup and stripping $^{11}$Li$+p$ scattering at $E=62$ MeV/nucleon, respectively. These
results give predictions because of missing experimental data for such processes accompanying the $^{11}$Li$+p$ scattering at $E\leq 100$ MeV/nucleon. For the
diffractive scattering we obtain values of the widths 98 MeV/c (for $n$=0) and 85 MeV/c (for $n$=1) and for the stripping reaction 79 MeV/c (for $n$=0) and 72 MeV/c
(for $n$=1), respectively, thus favoring the configuration in which the two valence neutrons occupy $1s$ state in $^{11}$Li. It is worth to be noted that the
calculated in our work widths for the $^{11}$Li breakup on the proton target are larger than those obtained in the experiments (around 50 MeV/c) for the reactions
of $^{11}$Li on the nuclear targets $^{9}$Be, $^{93}$Nb and $^{181}$Ta at energy $66$ MeV/nucleon \cite{Orr92} and on a wide range of targets ($^{9}$Be to
$^{238}$U) \cite{Orr95}. It is noted in \cite{Orr92,Orr95} that the width almost does not depend on the target's mass number and thus, it characterizes basically
the momentum distribution of two clusters. Our width for the stripping of $2n$-cluster is similar to the cases of $2n$ stripping from other nuclei (but not from
$^{11}$Li). It turns out that the account for the $2n$ binding in $^{11}$Li is not enough to obtain the observed widths in the scattering of $^{11}$Li on nuclei, as
well as on proton targets. We would like to mention also that we had a methodical task to calculate the widths using different wave functions ($n=0,1)$ of the
relative motion of the clusters. The results show similar values of the widths in both cases. Probably, it is difficult to solve the problem within our simplified
two-cluster model and thus, it must be considered in a more complicated three-body model. Also, obviously experiments on stripping and diffraction reactions of
$^{11}$Li on proton targets are highly desirable. This concerns measurements of the neutrons in the decay, as well.

\subsection{Single-particle density of $^{11}$Li in two-cluster
model}

In this subsection we would like to consider in more details the single-particle density distribution of $^{11}$Li, which can be calculated and applied instead of
phenomenological one in the analyses and interpretation of $^{11}$Li+$p$ experimental data. For this purpose, we adopt our cluster model, consisting of $^{9}$Li
core and halo $h=2n$. If one sets $\rho_{h}({\bf r}_{1})$ for the $h$-cluster and $\rho_{c}({\bf r}_{2})$ for $^{9}$Li nucleus, then the single-particle density
distribution of $^{11}$Li can be derived in analogy to Eq.~(\ref{eq:15d}) in the following form:
\begin{widetext}
\begin{eqnarray}
\rho(r)&=&\int d\phi \sin \theta d\theta \int ds s^{2} \left
[\rho_{h}({\bf r}_{h})
+\rho_{c}({\bf r}_{c})\right ]\rho_{0}^{(n)}({\bf s})=2\pi\int_{-1}^{1}dx \nonumber \\
&\times & \int_{0}^{\infty} ds s^{2} \left
[\rho_{h}\left(\sqrt{r^{2}-2(9/11)rsx+(9/11)^{2}s^{2}} \right
)+\rho_{c}\left(\sqrt{r^{2}+2(2/11)rsx+(2/11)^{2}s^{2}}\right
)\right ] \rho_{0}^{(n)}({\bf s}).
\label{eq:41}
\end{eqnarray}
\end{widetext}
The expression (\ref{eq:41}) indicates that the density of
$^{11}$Li can be calculated using the sum of the corresponding
densities of both clusters and folding it with the square of the
relative-motion wave function of the two clusters $|\phi_{00}({\bf
s})|^{2}$.

As a comment of our approach we would like to mention the
difference between the method to calculate the folding
$^{11}$Li$+p$ OP [Eq.~(\ref{eq:15d})] and that to estimate the
single-particle density of $^{11}$Li [Eq.~(\ref{eq:41})]. In fact,
in the former, the $U_{h}$ optical potential was not calculated as
a folding integral, but expressed through the $v_{np}$ potentials,
and therefore there we did not include the density of the $h=2n$
cluster. Instead, in Eq.~(\ref{eq:41}) we consider the $h$-cluster
density, together with the density of the $^{9}$Li core, both
being folded by the wave function of the relative motion of the
two clusters.

Further, in the calculations we use the LSSM density for the
$^{9}$Li cluster with rms radius $R_{c}$=2.31 fm
\cite{Karataglidis97} and for the $h$-halo we probe two densities:
the one being described by the Gaussian function (G density)
(e.g., \cite{Alkhazov2002})
\begin{equation}
\rho^{G}(r)=\left ( \frac{3}{2\pi R_{h}^{2}} \right
)^{3/2}\exp{\left (-\frac{3r^{2}}{2R_{h}^{2}} \right )}
\label{eq:42}
\end{equation}
and the other one is the symmetrized Fermi distribution (SF
density) (e.g., \cite{Burov77})
\begin{equation}
\rho^{SF}(r)=\rho_{0}\frac{\sinh{(R/a)}}{\cosh{(R/a)}+\cosh{(r/a)}},
\label{eq:43}
\end{equation}
where
\begin{equation}
\rho_{0}=\frac{3}{4\pi R^{3}}\left [1+\left (\frac{\pi a}{R}\right
)^{2}\right ]^{-1}
\label{eq:44}
\end{equation}
and the corresponding rms radius is:
\begin{equation}
<r^{2}>=R_{h}^{2}=\frac{3}{5}R^{2}\left [1+\frac{7}{3}\left
(\frac{\pi a}{R}\right )^{2}\right ].
\label{eq:45}
\end{equation}
The two densities [Eqs.~(\ref{eq:42}) and (\ref{eq:43})] are normalized to unity and substituting them in Eq.~(\ref{eq:41}) they have to be multiplied by a factor
of 2. As for the G density it has only one parameter, the rms radius of the halo $R_{h}$, that governs its behavior. First, we take $R_{h}$=2 fm which is almost
twice the nucleon radius. In principle, such a choice of $R_{h}$ is justified since the cluster inside the nucleus is "smeared" and, moreover, the folding procedure
itself (in which the relative motion function $\phi_{00}({\bf s})$ takes place with rms radius 4.93 fm, see also Sec.~III.A) ensures the $h$-cluster to be in the
periphery. Concerning the SF density, we perform calculations with a set of parameters $R$ and $a$, selected so that to obey rms $R_{h}$=2 fm (see the set SF1 in
Table~\ref{tab4}). For the choice of them the condition $R>\pi a$ must be satisfied and for a more convenience Eq.~(\ref{eq:45}) can be rewritten in the following
way:
\begin{equation}
R^{2}=\frac{5}{3}R_{h}^{2}-\frac{7}{3}(\pi a)^{2}.
\label{eq:46}
\end{equation}
\begin{table}
\caption{Values of the parameters of the symmetrized Fermi and
Gaussian density distributions, $h$- and $c$-cluster rms radii
$R_{h}$ and $R_{c}$, and deduced matter rms radii $R_{m}$ (in fm)
within the $^{9}$Li+2$n$ model of $^{11}$Li.} \label{tab4}
\begin{center}
\begin{tabular}{ccccccc}
\hline \hline \noalign{\smallskip}
Parametrization & & $R$ & $a$ & $R_{h}$ & $R_{c}$ & $R_{m}$ \\
\noalign{\smallskip}\hline\noalign{\smallskip}
SF1         & & 2.234 & 0.27 & 2  & $2.31^{a}$ & 2.77 ($n$=0) \\
            & &       &      &    &            & 2.93 ($n$=1) \\
G           & &       &      & 2  & $2.31^{a}$ & 2.77 ($n$=0) \\
            & &       &      &    &            & 2.93 ($n$=1) \\
SF2         & & 4.573 & 0.5  & 4  & $2.31^{a}$ & 3.32 ($n$=1) \\
0.2GG+0.8GO \protect\cite{Dobrovolsky2006} & &      &      & 5.98  & $2.52 $ & 3.42 \\
\noalign{\smallskip}\hline \hline $^{a}$ From LSSM for $^{9}$Li
\end{tabular}
\end{center}
\end{table}

\begin{figure}
\includegraphics[width=1.0\linewidth]{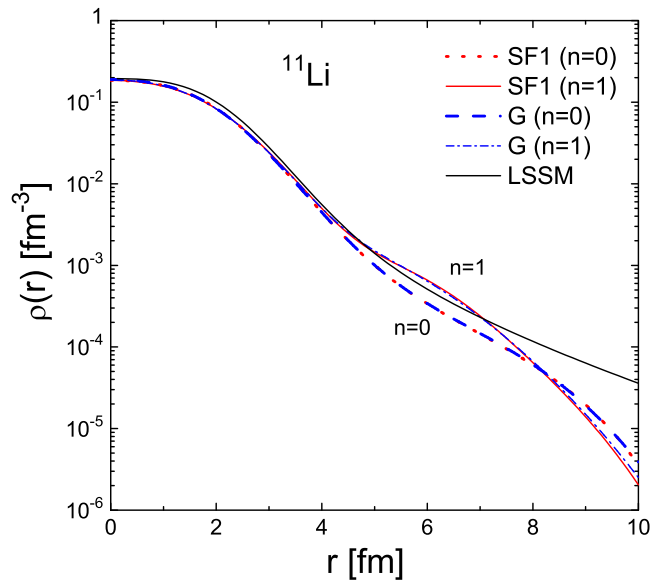}
\caption{(Color online) Single-particle density distribution of $^{11}$Li (normalized to $A=11$) obtained in the framework of the cluster model [Eq.~(\ref{eq:41})].
The $h$-cluster density distributions are taken in two forms: symmetrized Fermi distribution (SF1) and Gaussian function (G) with $R_{h}=2$ fm. The results are
presented for the cases of $n$=0 and $n$=1, respectively. The LSSM density is also given. \label{fig13}}
\end{figure}

The calculated single-particle density distributions of $^{11}$Li are presented in Fig.~\ref{fig13} together with the LSSM density. Results are shown for both $n$=0
and $n$=1 cases. As can be seen, the usage of two kinds of $h$-density SF1 and G yields very similar $^{11}$Li densities shown as the pair of the dotted and dashed
curves for $n$=0, and also as the solid and dot-dashed curves for $n$=1, correspondingly, in the whole region of $r$ up to 10 fm. In addition, all these four curves
are close at $r<4$ fm. However, the difference between the $n$=0 and $n$=1 pairs is seen in the interval $5<r<7$ fm, where the $n$=1 curves exhibit a "bump", while
the $n$=0 ones go down as compared to the case of the LSSM density of $^{11}$Li. Moreover, we note that the $^{11}$Li rms radius of 2.93 fm for $n$=1 curves is very
close to the LSSM value of 2.94 fm. The tail of the LSSM density is higher at $r>8$ fm than those of the cluster curves with $R_{h}=2$ fm, but as it was pointed out
in Ref.~\cite{Dobrovolsky2006} the calculated differential cross sections of $^{11}$Li$+p$ scattering are not sensitive to a possible long density tail at the
nuclear far periphery.

\begin{figure}
\includegraphics[width=1.0\linewidth]{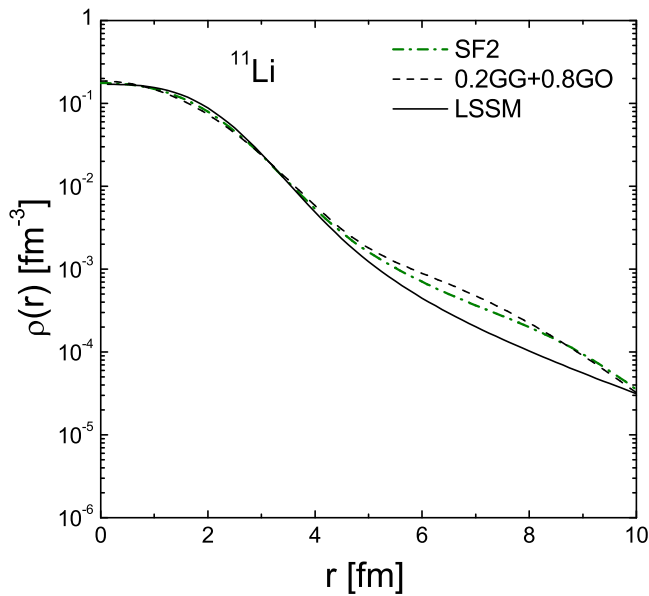}
\caption{(Color online) Single-particle density distribution of $^{11}$Li (normalized to $A=11$) obtained in the framework of the cluster model [Eq.~(\ref{eq:41})]
with symmetrized Fermi (SF2) distribution for the $h$-cluster density with $R_{h}=4$ fm (green dash-dotted line). The black dashed line represents the best density
parametrization that describes the $^{11}$Li$+p$ elastic scattering data \cite{Dobrovolsky2006}, while the black solid line is the LSSM density. \label{fig14}}
\end{figure}

The very pronounced halo nature of $^{11}$Li nucleus is mainly
supported by its large matter radius exhibited by Tanihata {\it et
al.} in Ref.~\cite{Tanihata85a}. Recently, a successful attempt to
get a "realistic" density of this nucleus was realized in
Ref.~\cite{Dobrovolsky2006}. In the latter the experimental data
at about 700 MeV/nucleon were described using the phenomenological
constituent cluster model of the $^{11}$Li density composed of two
terms 0.2GG+0.8GO with the Gaussian GG and the harmonic oscillator
GO functions together. The fitting procedure led to the total rms
matter radius $R_{m}=3.42$ fm of the whole density, where the
fitted values $R_{c}=2.52$ fm and $R_{h}=5.98$ fm of its separate
terms were interpreted as the core and $h$-halo radii,
respectively. These radii satisfy the relation
\begin{equation}
R_{m}^{2}=\frac{A_{c}R_{c}^{2}+A_{h}R_{h}^{2}}{A}, \;\;\;  A=A_{c}+A_{h},
\label{eq:47}
\end{equation}
($A_{c}$, $A_{h}$, and $A$ being number of nucleons in the core, in the $2n$-cluster and the nucleus, respectively) that is valid for the constituent model.
However, instead we may argue that the $^{9}$Li and $h$-systems can be considered as the true clusters only when in a cluster model they are folded [see
Eq.~(\ref{eq:41})] with the probability density of their relative motion. In Fig.~\ref{fig14} our result is shown as the SF2 curve when the value of $R_{h}=4$ fm is
taken to be twice larger than $R_{h}=2$ fm in the SF1 case. Also in the same figure we present the phenomenological 0.2GG+0.8GO density from
Ref.~\cite{Dobrovolsky2006}. Our SF2 parametrization leads to a value for the matter rms radius $R_{m}=3.32$ fm that is close to $R_{m}=3.42$ fm for the
phenomenological constituent model mentioned above. Thus, our folding method to calculate the single-particle density distribution [Eq.~(\ref{eq:41})] which takes
into account the relative motion of the clusters makes it possible to get realistic densities within cluster models without use of phenomenology. It is seen from
our analysis with SF2 parametrization that the $h$-cluster is really "smeared" in $^{11}$Li nucleus ($R_{h}=4$ fm) and that the averaging on a relative motion of
both clusters (which strongly depends on the $h$-cluster separation energy) plays an important role. This fact is confirmed also in Ref.~\cite{Dobrovolsky2006},
where the deduced halo radius $R_{h}=5.98$ fm is larger than the core radius $R_{c}=2.52$ fm by a factor of more than 2. However, the ambiguity remains in the
choice of the "best" density distribution of $^{11}$Li because only the $^{11}$Li$+p$ elastic scattering data are not sufficient.

\section{Conclusions}

The results of the present work can be summarized as follows:

(i) In the first part of the work (Sec.~II) the microscopic optical potentials and cross sections of $^{11}$Li$+p$ elastic scattering were calculated at the
energies of 62, 68.4, and 75 MeV/nucleon and were compared with the available experimental data. The direct ($V^{D}$) and exchange ($V^{EX}$) parts of the real OP
($V^{F}$) were calculated using the folding procedure with density-dependent M3Y (CDM3Y6-type) effective interaction based on the Paris $NN$ potential. The
imaginary part of OP ($W^{H}$) was calculated microscopically within the folding model based on the high-energy approximation. The LSSM densities
\cite{Karataglidis97} of protons and neutrons with exponential asymptotic behavior of $^{11}$Li that is the correct one were used in the calculations. The
spin-orbit contribution to the OP was also included in the calculations. The $^{11}$Li$+p$ elastic scattering cross sections and the total reaction cross sections
were calculated using the program DWUCK4 \cite{DWUCK}.

(ii) We pointed out that the regularization of our microscopic OP's is achieved by introducing the fitting parameters $N_R$, $N_I$, $N_R^{SO}$, $N_I^{SO}$ related
to the "depths" of the separate parts of OP. They are, in principle, the only free parameters of our approach, in contrast to other phenomenological ones and serve
as a quantitative test of the latter, i.e. the proximity of $N$'s values to unity shows the closeness of the approach to the reality. However, here the "ill-posed"
problem takes place because the fitting procedure is applied to a limited number of experimental data. The problem of the ambiguity of the $N$'s parameters have
been considered in our previous works \cite{Lukyanov2009,Lukyanov2010}. We used in the present work a physical constraint on the choice of the values of the $N$'s
parameters, namely the known behavior of the volume integrals $J_V$ and $J_W$ as functions of the incident energy for $E\leq 100$ MeV/nucleon \cite{Romanovsky}. We
compare the behavior of the values of $J_V$ and $J_W$ obtained in our work with those in the semi-phenomenological approach in Ref.~\cite{Hassan2009}, where much
more parameters have been used than in our microscopic method. We discuss in more details the problem arising from the behavior of $J_W$ at $E=62$ MeV/nucleon and
relate it to the quality of the data at larger angles ($\theta_{c.m.}>46^\circ$). We note that this problem had appeared also in \cite{Hassan2009}. Finally, we
obtained a definite set of the fitted $N$'s parameters that give satisfactory agreement of our results with the data of elastic $^{11}$Li$+p$ scattering cross
section using the physical criterion of the behavior of the volume integrals as functions of the energy.

(iii) We would like to mention that the values of the total cross
sections of scattering and reaction can serve as another physical
criterion for the $N$'s values. However, the corresponding
experimental data for these values are missing at the energy
interval considered in our work, so they are highly desirable.

(iv) As in our previous works \cite{Lukyanov2009,Lukyanov2010}, we would like to emphasize that a more successful explanation of the cross section data could be
given by accounting for virtual excitations of inelastic and decay channels of the reaction. For this reason, in Sec.~III of the present paper, apart from the usual
folding model based on the LSSM, we consider another folding approach that includes $^{11}$Li breakup suggesting a simple $^{9}$Li+2$n$ cluster model for its
structure. Both LSSM and cluster models of $^{11}$Li are capable to reproduce fairly well the two-neutron separation energy from $^{11}$Li. In Sec.~III we use the
procedure from the first part of our work (Sec.~II) for microscopic calculations of the necessary OP's in the breakup model for estimations of the elastic
scattering cross sections, as well as of the momentum distributions in the processes of the proton scattering on clusters and the corresponding $S$-functions in
$^{9}$Li$+p$ and $h$+$p$ scattering. The folding OP's calculated in the two parts of our work behave rather closely if one fits their strengths to the same elastic
scattering data, as it is done for $^{11}$Li$+p$ at energy 62 MeV/nucleon. Thus, the analysis of other types of the reaction mechanism, such as the $^{11}$Li
breakup, makes it possible to understand their significant role in the formation of the OP responsible for the $^{11}$Li$+p$ elastic scattering. It turns out that
the breakup channel gives $\sigma_{bu}^{tot}$ that exceeds 80\% from $\sigma_{R}^{tot}$, while it is around a half of $\sigma_{R}^{tot}$ in the case of
$^{6}$He+$^{12}$C (as obtained in Ref.~\cite{Lukyanov2011}).

(v) In the present work we give also predictions for the longitudinal momentum distributions of $^{9}$Li fragments produced in the breakup of $^{11}$Li at 62
MeV/nucleon on a proton target. We calculated the diffraction and stripping (when the cluster $2n$ leaves the elastic channel) cross sections of the reaction of
$^{11}$Li on proton target at energy 62 MeV/nucleon. We note that our breakup gives the width of the peak between 70 and 80 MeV/c, while the widths of about 50
MeV/c are known from the reactions of $^{11}$Li on nuclear targets $^{9}$Be, $^{93}$Nb and $^{181}$Ta at energy 66 MeV/nucleon. In relation with this, here we
should mention that at the energy of the range 60-70 MeV/nucleon a distortion due to the nuclear and Coulomb forces could affect the cross sections. We have in mind
also that our simplified two-cluster model could not give the correct answer and that it can be found in a more complicated three-body approach. Hence, this problem
remains open and requires further analysis. We emphasize the necessity of experiments on stripping and diffraction reactions of $^{11}$Li on proton targets at
energy $E<100$ MeV/nucleon.

(vi) We present results for the single-particle density
distribution of $^{11}$Li in the framework of a cluster model. Our
calculated density is close to the phenomenological one obtained
in Ref.~\cite{Dobrovolsky2006} by fitting to the experimental
differential cross sections of scattering of $^{11}$Li at 700
MeV/nucleon on a proton target. From a physical point of view the
cluster model allows more clear interpretation of the experimental
data and together with the phenomenological densities can be
applied as a pattern density to fit the data. Future measurements
of the cross sections for proton elastic scattering and momentum
distribution of the $^{9}$Li fragments in the $^{11}$Li breakup
reactions might provide supplemental information on the internal
spatial structure of the $^{11}$Li nucleus.

\begin{acknowledgments}
The authors are grateful to Professor N.S. Zelenskaya and
Professor S.N. Ershov for helpful discussions. The work is partly
supported by the Project from the Agreement for co-operation
between the INRNE-BAS (Sofia) and JINR (Dubna). Four of the
authors (D.N.K., A.N.A., M.K.G. and K.S.) are grateful for the
support of the Bulgarian Science Fund under Contract No.~02--285
and one of them (D.N.K.) under Contract
No.~DID--02/16--17.12.2009. The authors E.V.Z. and K.V.L. thank
the Russian Foundation for Basic Research (Grants Nos. 12-01-00396
and 13-01-00060) for partial support. K.S. acknowledges the
support of the Project BG-051P0001-3306-003.
\end{acknowledgments}


\begin{thebibliography}{99}

\bibitem{Tanihata85a} I. Tanihata, H. Hamagaki, O. Hashimoto, S.
Nagamiya, Y. Shida, N. Yoshikawa, O. Yamakawa, K. Sugimoto, T.
Kobayashi, D. E. Greiner, N. Takahashi, and Y. Nojiri, Phys. Lett. B
{\bf 160}, 380 (1985).

\bibitem{Tanihata85b} I. Tanihata, H. Hamagaki, O. Hashimoto, Y.
Shida, N. Yoshikawa, K. Sugimoto, O. Yamakawa, T. Kobayashi, and N.
Takahashi, Phys. Rev. Lett. {\bf 55}, 2676 (1985).

\bibitem{Tanihata88} I. Tanihata, T. Kobayashi, O. Yamakawa, T.
Shimoura, K. Ekuni, K. Sugimoto, N. Takahashi, T. Shimoda, and H.
Sato, Phys. Lett. B {\bf 206}, 592 (1988).

\bibitem{Mittig87} W. Mittig, J. M. Chouvel, Z. W. Long, L. Bianchi,
A. Cunsolo, B. Fernandez, A. Foti, J. Gastebois, A. Gillibert, C.
Gregoire, Y. Schutz, and C. Stephan, Phys. Rev. Lett. {\bf 59}, 1889
(1987).

\bibitem{Hansen87} P. G. Hansen and B. Jonson, Eur. Lett. {\bf 4}, 409 (1987).

\bibitem{Migdal73} A. B. Migdal, Sov. J. Nucl. Phys. {\bf 16}, 238
(1973).

\bibitem{Kobayashi88} T. Kobayashi {\it et al.}, Phys. Rev. Lett. {\bf
60}, 2599 (1988).

\bibitem{Bertsch89} G. F. Bertsch, B. A. Brown, and H. Sagawa, Phys.
Rev. C {\bf 39}, 1154 (1989); T. Hoshino, H. Sagawa, and A. Arima,
Nucl. Phys. A {\bf 506}, 271 (1990); L. Johannsen, A. S. Jensen,
and P. G. Hansen, Phys. Lett. B {\bf 244}, 357 (1990); A. C.
Hayes, {\it ibid} {\bf 254}, 15 (1991); G. F. Bertsch and H.
Esbensen, Ann. Phys. {\bf 209}, 327 (1991); Y. Tosaka and Y.
Suzuki, Nucl. Phys. A {\bf 512}, 46 (1990).

\bibitem{Anne90} R. Anne {\it et al.}, Phys. Lett. B {\bf 250}, 19
(1990).

\bibitem{Esbensen91} H. Esbensen, Phys. Rev.
C {\bf 44}, 440 (1991); {\it ibid.} {\bf 53}, 2007 (1996).

\bibitem{Orr92} N. A. Orr, N. Anantaraman, Sam M. Austin, C.A. Bertulani, K. Hanold,
J. H. Kelley, D. J. Morrissey, B. M. Sherrill, G. A. Souliotis, M. Thoennessen, J. S. Winfield
and J. A. Winger, Phys. Rev. Lett. {\bf 69}, 2050 (1992).

\bibitem{Orr95} N. A. Orr, N. Anantaraman, Sam M. Austin, C.A. Bertulani, K. Hanold,
J. H. Kelley, R. A. Kryger, D. J. Morrissey, B. M. Sherrill, G. A. Souliotis, M. Steiner, M. Thoennessen, J. S. Winfield, J. A. Winger, and B. M. Young, Phys. Rev.
C {\bf 51}, 3116 (1995).

\bibitem{Baye2010} D. Baye and P. Capel, {\it Lecture Notes in Physics},
vol. 848, pp.121-163 (2012).

\bibitem{Barranco96} F. Barranco, E. Vigezzi, and R. A. Broglia, Z. Phys. A
{\bf 356}, 45 (1996).

\bibitem{Hencken96} K. Hencken, G. Bertsch, and H. Esbensen, Phys. Rev. C {\bf 54},
3043 (1996).

\bibitem{Bertulani2004} C. A. Bertulani and P. G. Hansen, Phys.
Rev. C {\bf 70}, 034609 (2004).

\bibitem{Bertulani92} C. A. Bertulani and K. W. McVoy, Phys.
Rev. C {\bf 46}, 2638 (1992).

\bibitem{Ershov2004} S. N. Ershov, B. V. Danilin, J. S. Vaagen, A. A. Korsheninnikov,
and I. J. Thompson, Phys. Rev. C {\bf 70}, 054608 (2004).

\bibitem{Bertulani2002} C. A. Bertulani, M. Hussein, and G. Muenzenberg,
{\it Physics of Radioactive Beams} (Nova Science, Hauppage, New York, 2002), ISBN: 1-59033-141-9.

\bibitem{Dobrovolsky2006} A. V. Dobrovolsky, G. D. Alkhazov, M. N. Andronenko,
A. Bauchet, P. Egelhof, S. Fritz, H. Geissel, C. Gross, A. V.
Khanzadeev, G. A. Korolev, G. Kraus, A. A. Lobodenko, G.
M\"{u}nzenberg, M. Mutterer, S. R. Neumaier, T. Sch\"{a}fer, C.
Scheidenberger, D. M. Seliverstov, N. A. Timofeev, A. A. Vorobyov,
and V. I. Yatsoura, Nucl. Phys. A {\bf 766}, 1 (2006).

\bibitem{Moon92} C.-B. Moon, M. Fujimaki, S. Hirenzaki, N. Inabe, K.
Katori, J. C. Kim, Y. K. Kim, T. Kobayashi, T. Kubo, H. Kumagai,
S. Shimoura, T. Suzuki, and I. Tanihata, Phys. Lett. B {\bf 297},
39 (1992).

\bibitem{Korsh97c} A. A. Korsheninnikov, E. A. Kuzmin, E. Yu.
Nikolskii, O. V. Bochkarev, S. Fukuda, S. A. Goncharov, S. Ito, T.
Kobayashi, S. Momota, B. G. Novatskii, A. A. Ogloblin, A. Ozawa,
V. Pribora, I. Tanihata, and K. Yoshida, Phys. Rev. Lett. {\bf
78}, 2317 (1997).

\bibitem{Korsh96} A. A. Korsheninnikov, E. Yu. Nikolskii, T.
Kobayashi, A. Ozawa, S. Fukuda, E. A. Kuzmin, S. Momota, B. G.
Novatskii, A. A. Ogloblin, V. Pribora, I. Tanihata, and K.
Yoshida, Phys. Rev. C {\bf 53}, R537 (1996).

\bibitem{Zhukov93} M. V. Zhukov, B. V. Danilin, D. V. Fedorov, J. M.
Bang, I. J. Thompson, and J. S. Vaagen, Phys. Rep. {\bf 231}, 151
(1993).

\bibitem{Suzuki93} Y. Suzuki, K. Yabana, and Y. Ogawa, Phys. Rev. C
{\bf 47}, 1317 (1993).

\bibitem{Kohno93} M. Kohno, Phys. Rev. C {\bf 48}, 3122 (1993).

\bibitem{Chaudhuri94} A. K. Chaudhuri, Phys. Rev. C {\bf 49}, 1603 (1994).

\bibitem{Kanungo97} R. Kanungo and C. Samanta, Nucl. Phys. A {\bf
617}, 265 (1997).

\bibitem{Kim2001} Y. J. Kim and M. H. Cha, Int. J. Mod. Phys. E {\bf 10},
91 (2001).

\bibitem{Crespo96} R. Crespo, J. A. Tostevin, and I. J. Thompson, Phys. Rev. C
{\bf 54}, 1867 (1996).

\bibitem{Amos2005} K. Amos, W. A. Richter, S.Karataglidis, and B. A.
Brown, Phys. Rev. Lett. {\bf 96}, 032503 (2006); P. K. Deb, B. C.
Clark, S. Hama, K. Amos, S. Karataglidis, and E. D. Cooper, Phys.
Rev. C {\bf 72}, 014608 (2005).

\bibitem{Avrigeanu2000} M. Avrigeanu, G. S. Anagnostatos, A. N.
Antonov, and J. Giapitzakis, Phys. Rev. C {\bf 62}, 017001 (2000);
M. Avrigeanu, G. S. Anagnostatos, A. N. Antonov, and V. Avrigeanu,
Int. J. Mod. Phys. E {\bf 11}, 249 (2002); M. Avrigeanu, A. N.
Antonov, H. Lenske, and I. Stetcu, Nucl. Phys. A {\bf 693}, 616
(2001).

\bibitem{Satchler79} G. R. Satchler and W. G. Love, Phys. Rep. {\bf
55}, 183 (1979); G. R. Satchler, {\it Direct Nuclear Reactions}
(Clarendon, Oxford, 1983).

\bibitem{Khoa1993} D. T. Khoa and W. von Oertzen, Phys. Lett. B {\bf
304}, 8 (1993); {\bf 342}, 6 (1995); D. T. Khoa, W. von Oertzen,
and H. G. Bohlen, Phys. Rev. C {\bf 49}, 1652 (1994); D. T. Khoa,
W. von Oertzen and A. A. Ogloblin, Nucl. Phys. {\bf A602}, 98
(1996); Dao T. Khoa and Hoang Sy Than, Phys. Rev. C {\bf 71},
014601 (2005); O. M. Knyaz'kov, Sov. J. Part. Nucl. {\bf 17}, 137
(1986).

\bibitem{Khoa2000} D. T. Khoa and G. R. Satchler, Nucl. Phys. A {\bf
668}, 3 (2000).

\bibitem{Khoa97} D. T. Khoa, G. R. Satchler, and W. von Oertzen, Phys.
Rev. C {\bf 56}, 954 (1997).

\bibitem{Hassan2009} M. Y. M. Hassan, M. Y. H. Farag, E. H. Esmael,
and H. M. Maridi, Phys. Rev. C {\bf 79}, 014612 (2009).

\bibitem{Farag2012} M. Y. H. Farag, E. H. Esmael, and H. M.
Maridi, Eur. Phys. J. A {\bf 48}, 154 (2012).

\bibitem{Lukyanov2004a} K. V. Lukyanov, E. V. Zemlyanaya, and V. K.
Lukyanov, JINR Preprint P4-2004-115, Dubna, 2004; Phys. At. Nucl.
{\bf 69}, 240 (2006).

\bibitem{Shukla2003} P. Shukla, Phys. Rev. C {\bf 67}, 054607 (2003).

\bibitem{Glauber} R. J. Glauber, {\it Lectures in Theoretical Physics}
(New York, Interscience, 1959), p.315.

\bibitem{Sitenko} A. G. Sitenko, Ukr. Fiz. J. {\bf 4}, 152 (1959).

\bibitem{Karataglidis97} S. Karataglidis, P. G. Hansen, B. A. Brown,
K. Amos, and P. J. Dortmans, Phys. Rev. Lett. {\bf 79}, 1447
(1997); S. Karataglidis, P. J. Dortmans, K. Amos, and C. Bennhold,
Phys. Rev. C {\bf 61}, 024319 (2000).

\bibitem{Lukyanov2007} K. V. Lukyanov, V. K. Lukyanov, E. V.
Zemlyanaya, A. N. Antonov, and M. K. Gaidarov, Eur. Phys. J. A
{\bf 33}, 389 (2007).

\bibitem{Lukyanov2009} V. K. Lukyanov, E. V. Zemlyanaya, K. V.
Lukyanov, D. N. Kadrev, A. N. Antonov, M. K. Gaidarov, and S. E.
Massen, Phys. Rev. C {\bf 80}, 024609 (2009).

\bibitem{Lukyanov2010} V. K. Lukyanov, D. N. Kadrev, E. V. Zemlyanaya,
A. N. Antonov, K. V. Lukyanov, and M. K. Gaidarov, Phys. Rev. C
{\bf 82}, 024604 (2010).

\bibitem{Lukyanov2012} V. K. Lukyanov, D. N. Kadrev, E. V. Zemlyanaya, A. N. Antonov,
K. V. Lukyanov, M. K. Gaidarov, and K. Spasova, Phys. At. Nucl. {\bf 75}, 1407 (2012).

\bibitem{DWUCK} P. D. Kunz and E. Rost, in {\it Computational Nuclear
Physics}, edited by K. Langanke {\it et al.} (Springer-Verlag, New York, 1993), Vol.2, p.88.
\bibitem{Shukla2001} P. Shukla, arXiv: nucl-th/0112039.

\bibitem{Charagi92} S. Charagi and G. Gupta, Phys. Rev. C {\bf 41},
1610 (1990); {\bf 46}, 1982 (1992).

\bibitem{Xiangzhow98} C. Xiangzhow, F. Jun, S. Wenqing, M. Yugang, W.
Jiansong, and Y. Wei, Phys. Rev. C {\bf 58}, 572 (1998).

\bibitem{Tikhonov77} A. N. Tikhonov and V. Y. Arsenin, {\it Solutions
of Ill-Posed Problems}, (V. H. Winston and Sons, Wiley, New York, 1977).

\bibitem{Romanovsky} E. A. Romanovsky {\it et al.}, Bulletin of the
Russian Academy of Sciences: Physics {\bf 62}, No.1, 150 (1998).

\bibitem{Lukyanov2011} V. K. Lukyanov, E. V. Zemlyanaya, and K. V.
Lukyanov, Int. J. Mod. Phys. E {\bf 20}, 2039 (2011).

\bibitem{Thompson94} I. J. Thompson and M. V. Zhukov, Phys. Rev. C {\bf 49},
1904 (1994).

\bibitem{Zelenskaya} L. I. Galanina and N. S. Zelenskaya, Bulletin of the
Russian Academy of Sciences: Physics {\bf 77}, No.4, 383 (2013); Phys. At. Nucl. {\bf 76}, No.12 (2013), in print.

\bibitem{Myo2007} T. Myo, K. Kato, H. Toki, and K. Ikeda, Phys. Rev. C {\bf 76},
024305 (2007).

\bibitem{Tostevin98} J. A. Tostevin, R. C. Johnson, and J. S.
Al-Khalili, Nucl. Phys. A {\bf 630}, 340c (1998).

\bibitem{Ershov2010} S. N. Ershov, L. V. Grigorenko, J. S. Vaagen,
and M. V. Zhukov, J. Phys. G {\bf 37}, 064026 (2010).

\bibitem{Thompson77} D. R. Thompson, M. LeMere, and Y. C. Tang, Nucl. Phys. A {\bf 286}, 53
(1977); Y. C. Tang, M. LeMere, and D. R. Thompson, Phys. Rep. {\bf 47}, 167 (1978).

\bibitem{Alkhazov2002} G. D. Alkhazov, A. V. Dobrovolsky, P. Egelhof, H. Geissel,
H. Irnich, A. V. Khanzadeev, G. A. Korolev, A. A. Lobodenko, G.
M\"{u}nzenberg, M. Mutterer, S. R. Neumaier, W. Schwab, D. M.
Seliverstov, T. Suzuki, and A. A. Vorobyov, Nucl. Phys. A {\bf
712}, 269 (2002).

\bibitem{Burov77} V. V. Burov and V. K. Lukyanov, Preprint JINR,
R4-11098, 1977, Dubna; V. V. Burov, D. N. Kadrev, V. K. Lukyanov,
and Yu. S. Pol', Phys. At. Nuclei {\bf 61}, 525 (1998).

\end{thebibliography}
\end{document}